\documentclass[aps, prd, floatfix, nofootinbib, superscriptaddress, twocolumn]{revtex4-1}

\usepackage{latexsym}
\usepackage{amsmath}
\usepackage{amssymb}
\usepackage{amsfonts}

\usepackage[mathscr,scaled=1.15]{urwchancal}
\DeclareFontFamily{OT1}{pzc}{}
\DeclareFontShape{OT1}{pzc}{m}{it}%
{<-> s * [1.15] pzcmi7t}{}
\DeclareMathAlphabet{\mathpzc}{OT1}{pzc}{m}{it}

\usepackage{color}

\usepackage{supertabular}
\usepackage{placeins}
\usepackage{epsfig}
\usepackage{graphicx}

\definecolor{purple}{rgb}{0.5,0,0.5}
\definecolor{blue}{rgb}{0.0,0,0.9}
\definecolor{prdblue}{rgb}{0.133,0.118,0.498}
\usepackage[colorlinks=true, pdfstartview=FitV, linkcolor=prdblue, citecolor= prdblue, urlcolor=prdblue]{hyperref}

\begin{document}

% Use the \preprint command to place your local institutional report number in the upper righthand corner of the title
% page in preprint mode. Multiple \preprint commands are allowed. Use the 'preprintnumbers' class option to override
% journal defaults to display numbers if necessary
%\preprint{}

%\title{$\gamma^{\ast}N\to N(1440)\,1/2^+$ electromagnetic transition form factors}
%\title{Proton-to-$\Delta(1600)$ transition form factors.}
\title{Transition form factors: $\mathbf{\gamma^\ast + p \to \Delta(1232)}$, $\mathbf{\Delta(1600)}$ }

\author{Y. Lu}
\email[]{luya@nju.edu.cn}
\affiliation{Department of Physics, Nanjing University, Nanjing, Jiangsu 210093, China}

\author{C. Chen}
\email[]{chenchen@ift.unesp.br}
\affiliation{Instituto de F\'isica Te\'orica, Universidade Estadual Paulista, Rua Dr.~Bento Teobaldo Ferraz, 271, 01140-070 S\~ao Paulo, SP, Brazil}

\author{Z.-F. Cui}
\email[]{phycui@nju.edu.cn}
\affiliation{Department of Physics, Nanjing University, Nanjing, Jiangsu 210093, China}

\author{C. D. Roberts}
\email[]{cdroberts@anl.gov}
%\thanks{}
\affiliation{Physics Division, Argonne National Laboratory, Lemont, Illinois
60439, USA}

\author{S. M. Schmidt}
\email[]{s.schmidt@fz-juelich.de}
\affiliation{
Institute for Advanced Simulation, Forschungszentrum J\"ulich and JARA, D-52425 J\"ulich, Germany}

\author{J. Segovia}
\email[]{jsegovia@upo.es}
\affiliation{Departamento de Sistemas F\'{\i}sicos, Qu\'{\i}micos y Naturales,
Universidad Pablo de Olavide, E-41013 Sevilla, Spain}

%\affiliation{Department of Physical, Chemical and Natural Systems,
%University of Pablo de Olavide, Ctra.\ de Utrera, km 1,
%E-41013 Seville, Spain}

%\author{Shu-Sheng Xu}
%\email[]{xuss@nju.edu.cn}
%\affiliation{College of Science, Nanjing University of Posts and Telecommunications, Nanjing 210023, China}

\author{H.-S. Zong}
\email[]{zonghs@nju.edu.cn}
\affiliation{Department of Physics, Nanjing University, Nanjing, Jiangsu 210093, China}
\affiliation{Joint Center for Particle, Nuclear Physics and Cosmology, Nanjing, Jiangsu 210093, China}


\date{05 April 2019}
%\date{11 March 2019}

\begin{abstract}
Electroproduction form factors describing the $\gamma^\ast p \to \Delta^+(1232), \Delta^+(1600)$ transitions are computed using a fully-dynamical diquark-quark approximation to the Poincar\'e-covariant three-body bound-state problem in relativistic quantum field theory.  In this approach, the $\Delta(1600)$ is an analogue of the Roper resonance in the nucleon sector, appearing as the simplest radial excitation of the $\Delta(1232)$.  Precise measurements of the $\gamma^\ast p \to \Delta^+(1232)$ transition already exist on $0 \leq Q^2 \lesssim 8\,$GeV$^2$ and the calculated results compare favourably with the data outside the meson-cloud domain.  The predictions for the $\gamma^\ast p \to \Delta^+(1600)$ magnetic dipole and electric quadrupole transition form factors are consistent with the empirical values at the real photon point, and extend to $Q^2 \approx 6 m_p^2$, enabling a meaningful direct comparison with experiment once analysis of existing data is completed.  In both cases, the electric quadrupole form factor is particularly sensitive to deformation of the $\Delta$-baryons.  Interestingly, whilst the $\gamma^\ast p \to \Delta^+(1232)$ transition form factors are larger in magnitude than those for $\gamma^\ast p \to \Delta^+(1600)$ in some neighbourhood of the real photon point, this ordering is reversed on $Q^2 \gtrsim 2 m_p^2$, suggesting that the $\gamma^\ast p \to \Delta^+(1600)$ transition is more localised in configuration space.
\end{abstract}

\maketitle

%%%%%%%%%%%%%%%%%%%%%%%%%%%%%%%%%%%%%%%%%%%%%%%%%%%%%%%%%%%%%%%%%%%%%%%%%%%%%%%%%%%%%%%%%%%%%%%%%%%%%%%%%%%%%%%%%%%%%%%
% 4500 words

%\noindent\textbf{1.$\;$Introduction}.
\section{Introduction}
The $\Delta(1232)$ family of baryons were the first resonances discovered in $\pi N$ reactions \cite{Fermi:1952zz, Anderson:1952nw, Nagle:1984sg}.  With positive parity, isospin $I=\frac{3}{2}$, total-spin $J=\frac{3}{2}$ and no net strangeness \cite{Tanabashi:2018oca}, the $\Delta^{+,0}$ members of this quadruplet have conventionally been viewed as the lightest isospin- and spin-flip excitations of the proton and neutron, respectively.  Hence, since protons and neutrons (nucleons, $N$) are the basic elements of all nuclei, developing a detailed understanding of the $\Delta$-baryons is of fundamental importance.  Without this, hadron physics remains at a level akin to atomic physics based only on knowledge of the hydrogen atom's ground state.

Given that pions are a complex probe, there are advantages in exploiting the relative simplicity of virtual photons in order to chart $\Delta$-resonance structure.  Elastic form factors are empirically inaccessible because the $\Delta(1232)$-baryon lifetime is too small: $\tau_\Delta \sim 10^{-26}\,\tau_n$, where $\tau_n$ is the lifetime of a free neutron \cite{Tanabashi:2018oca}.  On the other hand, by exploiting intense, energetic electron-beams at the Thomas Jefferson National Accelerator Facility, $\gamma^\ast p \to \Delta^+$ data are now available for $0 \leq Q^2 \lesssim 8\,$GeV$^2$ \cite{Aznauryan:2011ub, Aznauryan:2011qj, Aznauryan:2012ba}.
These data have stimulated much theoretical analysis and speculation about, \emph{inter alia}:
the relevance of perturbative QCD (pQCD) to processes involving moderate momentum transfers \cite{Carlson:1985mm, Pascalutsa:2006up, Aznauryan:2011qj, Aznauryan:2012ba, Eichmann:2011aa, Segovia:2013rca, Segovia:2013uga, Segovia:2014aza, Segovia:2016zyc};
hadron shape deformation \cite{Eichmann:2011aa, Alexandrou:2012da, Santopinto:2012nq, Segovia:2013rca, Segovia:2013uga, Segovia:2014aza, Segovia:2016zyc, Sanchis-Alepuz:2017mir, Buchmann:2018nmu};
and the role that resonance electroproduction experiments can play in exposing nonperturbative aspects of QCD, such as the nature of confinement and dynamical chiral symmetry breaking (DCSB) \cite{Aznauryan:2012ba, Roberts:2015dea, Burkert:2018oyl, Roberts:2018hpf}.
% -- two emergent phenomena that play a key role in forming the bulk of visible matter in the Universe \cite{national2012Nuclear}.

Just above the $\Delta$-baryon level lies the nucleon's first positive-parity excitation, \emph{i.e}.\ the Roper resonance, labelled $N(1440)\,1/2^+$.  Discovered in 1963 \cite{Roper:1964zza, BAREYRE1964137, AUVIL196476, PhysRevLett.13.555, PhysRev.138.B190}, its characteristics were long the source of puzzlement because, \emph{e.g}.\ constituent-quark potential models typically (and erroneously) produce a spectrum in which this excitation lies above the first negative-parity state $N(1535)\,1/2^-$ \cite{Capstick:2000qj, Crede:2013sze, Giannini:2015zia}.  This has now changed following: acquisition and analysis of high-precision proton-target exclusive electroproduction data with single- and double-pion final states, on a large energy domain and with momentum-transfers out to $Q^2\approx 5\,$GeV$^2$; development of a dynamical reaction theory capable of simultaneously describing all partial waves extracted from available, reliable data; and formulation and application of a Poincar\'e covariant approach to the continuum bound-state problem in relativistic quantum field theory.  Today, it is widely accepted that the Roper is, at heart, the first radial excitation of the nucleon, consisting of a well-defined dressed-quark core that is augmented by a meson cloud, which both reduces the Roper's core mass by approximately 20\% and contributes materially to the electroproduction form factors at low-$Q^2$ \cite{Golli:2017nid, Burkert:2017djo}.

A similar pattern of energy levels is found in the spectrum of $\Delta$-baryons.  Namely, contradicting quark-model predictions \cite{Capstick:2000qj, Crede:2013sze, Giannini:2015zia},
% table 15 p 69 in Crede:2013sze
the first positive-parity excitation, $\Delta(1600)\,3/2^+$, lies below the negative parity $\Delta(1700)\,3/2^-$, with the splitting being approximately the same as that in the nucleon sector.
This being the case and given the Roper-resonance example, it is likely that elucidating the nature of the $\Delta(1600)\,3/2^+$-baryon will require both (\emph{i}) data on its electroproduction form factors which extends well beyond the meson-cloud domain and (\emph{ii}) predictions for these form factors to compare with that data.
The data exist \cite{Trivedi:2018rgo, Burkert:2019opk}; and can be analysed with this aim understood.  Herein, therefore, we provide the theoretical predictions.

%Currently, I am completing the initial description of the best accuracy CLAS pi^+pi^-p electroproduction data (Ralf) in the mass range W<2.0 GeV and 2.0 GeV^2 < Q^2< 4.5 GeV^2. By the end of  19' I will have electrocouplings of all resonances in the mass range <1.65 GeV and 2.0 GeV^2 < Q^2< 4.5 GeV^2.

%Right now I just turn off \Delta(1600)3/2+. Reasonable data description was achieved in the asformentioned Q^2 range under this assumption, therefore, consistent to the data. Could you please give me the DSE results on  \Delta(1600)3/2+ electrocouplings. I will use them as the start point and fit to the 9 independent pi^+pi^-p electroproduction cross sections. I have no any other initial estimate.

%Spectrum information is not enough ... too easy to reproduce masses ... almost independent of the interaction ... even contact interaction is good enough.
%
%Like Roper, only high-quality electroproduction data can provide the answer.
%
%We provide predictions based on continuum QCD that will be tested with forthcoming CLAS12 data

Our treatment of the nucleon, $\Delta(1232)$- and $\Delta(1600)$-baryons, and the associated $\gamma N\to \Delta$ transitions is based on Refs.\,\cite{Segovia:2014aza, Chen:2019fzn}.   Capitalising on this tight connection, herein we only sketch the elements of our calculation.  (Isospin symmetry is assumed throughout.)  Moreover, with nothing changed, our study delivers a unification of the $N\to \Delta(1232)$ and $N\to \Delta(1600)$ transitions.

Section~\ref{SecFaddeev} explains the quark-diquark approximation to the baryon problem in the context of a Poincar\'e-covariant Faddeev equation and discusses the solutions obtained for the $\Delta(1232)$-baryon and its first positive-parity excitation.  The $\gamma^\ast p \to \Delta$ transition current and associated form factors are described in Sec.\,\ref{SecTransition}.  Section~\ref{SecTFFs1232} reports results for the $\gamma^\ast p\to \Delta(1232)$ transition, providing comparisons with data and other analyses.  The $\gamma^\ast p\to \Delta(1600)$ transition form factors are discussed in Sec.\,\ref{SecTFFs1600}; Sec.\,\ref{1600Dissections} describes their diquark and scatterer dissections; and Sec.\,\ref{epilogue} provides a summary and offers perspectives.

\section{Baryon Wave Functions}
\label{SecFaddeev}
\begin{figure}[t]
\centerline{%
\includegraphics[clip, width=0.45\textwidth]{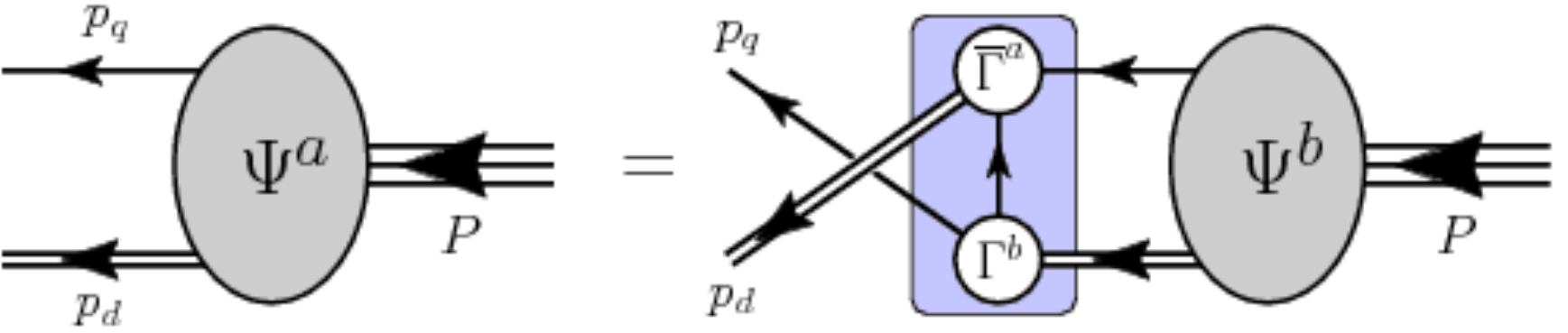}}
\caption{\label{figFaddeev}
Faddeev equation: a linear integral equation for the matrix-valued function $\Psi$, being the Faddeev amplitude for a baryon of total momentum $P= p_q + p_d$, which expresses the relative momentum correlation between the dressed-quarks and -nonpointlike-diquarks within the baryon.  The shaded rectangle demarcates the kernel of the Faddeev equation:
\emph{single line}, dressed-quark propagator; $\Gamma$,  diquark correlation amplitude; and \emph{double line}, diquark propagator. }
\end{figure}

%
%Our treatment of the nucleon, $\Delta(1232)$- and $\Delta(1600)$-baryons and the associated $N\to \Delta$ transitions is based on Refs.\,\cite{Segovia:2014aza, Chen:2019fzn}: nothing is changed; hence, we can make direct comparisons between the $N\to \Delta(1232)$ transition form factors computed therein and the $N\to \Delta(1600)$ form factors described below.  Capitalising on this tight connection, herein we only sketch the elements of our calculation.  (Isospin symmetry is assumed throughout.)
%
In relativistic quantum field theory, baryon structure is described by a Faddeev amplitude, obtained from a Poincar\'e-covariant Faddeev equation, which sums all possible quantum field theoretical exchanges and interactions that can take place between the three dressed-quarks that characterise its valence-quark content.
A dynamical prediction of Faddeev equation studies that employ realistic quark-quark interactions \cite{Binosi:2014aea, Binosi:2016wcx, Binosi:2016nme, Rodriguez-Quintero:2018wma} is the appearance of nonpointlike quark$+$quark (diquark) correlations within baryons, whose characteristics are determined by DCSB \cite{Cahill:1987qr, Maris:2002yu, Maris:2004bp}.  Consequently, the baryon bound-state problem is transformed into solving the linear, homogeneous matrix equation in Fig.\,\ref{figFaddeev} \cite{Cahill:1988dx, Burden:1988dt, Cahill:1988zi, Reinhardt:1989rw, Efimov:1990uz}.
Its key elements are the dressed-quark and -diquark propagators, and the diquark Bethe-Salpeter amplitudes.

Evidence supporting the presence of diquark correlations in baryons is accumulating, \emph{e.g}.\ Refs.\,\cite{Eichmann:2009qa, Cates:2011pz, Segovia:2013rca, Roberts:2013mja, Segovia:2014aza, Segovia:2015ufa, Eichmann:2016jqx, Segovia:2016zyc, Eichmann:2016hgl, Eichmann:2016nsu, Lu:2017cln, Chen:2017pse, Mezrag:2017znp, Roberts:2018hpf, Chen:2018nsg, Chen:2019fzn}.
It should be emphasised that these correlations are fully dynamical and appear in a Faddeev kernel which requires their continual breakup and reformation.  Consequently, they are vastly different from the static, pointlike diquarks introduced originally \cite{Anselmino:1992vg} in an attempt to solve the so-called ``missing resonance'' problem \cite{Aznauryan:2011ub}.  In fact, consistent with numerical simulations of lattice-regularised QCD \cite{Edwards:2011jj}, the spectrum of states produced by the Faddeev equation in Fig.\,\ref{figFaddeev} possesses a richness that cannot be explained by a two-body model.

To define the Faddeev equation in Fig.\,\ref{figFaddeev}, we employ the elements specified in Refs.\,\cite{Segovia:2014aza, Chen:2019fzn}, which provide a successful description of the spectrum and structure of octet and decuplet baryons and their positive-parity excitations, and are part of a body of work that unifies a large array of hadron properties \cite{Roberts:2015lja, Horn:2016rip, Eichmann:2016yit, Burkert:2017djo}.
A key to these successes is DCSB, which produces a dressed-quark mass-scale \cite{Binosi:2016wcx}: $M_D\simeq 0.4\,$GeV, whose value underlies the natural size for mass-dimensioned quantities in the light-quark sector of the Standard Model.

%Ms=0.794232 Ma=0.891548
With the inputs drawn from Refs.\,\cite{Segovia:2014aza, Chen:2019fzn} (including light-quark scalar and axial-vector diquark masses $m_{0^+}=0.79\,$GeV, $m_{1^+} = 0.89\,$GeV, respectively) one can readily construct the relevant Faddeev equation kernels and use \emph{ARPACK} software \cite{Arpack} to obtain the mass and Faddeev amplitude of the $(I,J^P)=(1/2,1/2^+)$ ground-state (proton) and the two lightest $(I,J^P)=(3/2,3/2^+)$ states, which we identify with the $\Delta(1232)$- and $\Delta(1600)$-baryons.  The masses are (in GeV):
\begin{equation}
\label{eqMasses}
%\textstyle
%\begin{array}{l|cc}
%            & \mbox{Nucleon\,(N)} & \mbox{Roper\,(R)} \\\hline
%\mbox{mass} & 1.18 & 1.73
%\end{array}\,.
\begin{array}{ccc}
m_p & m_{\Delta(1232)} & m_{\Delta(1600)} \\
1.19 & 1.35 & 1.79
\end{array}\,.
\end{equation}
These values correspond to the locations of the lowest-magnitude poles in the three-quark scattering problems in the given channels.

The residues associated with these poles are the Poincar\'e-covariant wave functions, $\chi(\ell^2,\ell\cdot P; P^2)$, where $\ell$ is the quark-diquark relative momentum.  For every baryon considered herein, eight scalar functions are required to completely describe the system, each associated with a particular Dirac-matrix structure.  For instance, the (amputated) Faddeev amplitude of any $(I,J^P)=(3/2,3/2^+)$ baryon can be written in the following form:
\begin{align}
\nonumber
 \psi^{\Delta}(p_i,\alpha_i,\sigma_i) & =  \sum_{d\in \Delta} \, [\Gamma^d_{1^+\mu}(k;K)]^{\alpha_1 \alpha_2}_{\sigma_1 \sigma_2} \, \\
  & \quad  \times \Delta^{1^+ d}_{\mu\nu}(K) \, [\varphi_{\nu\rho}^{\Delta d}(\ell;P) u_\rho(P)]^{\alpha_3}_{\sigma_3} \,,
\label{FaddeevAmp10}
\end{align}
where:
$(p_i,\sigma_i,\alpha_i)$ are the momentum, spin and isospin labels of the quarks constituting the bound state;
$P=p_1 + p_2 + p_3=p_d+p_q$ is the total momentum of the baryon;
$k=(p_1-p_2)/2$, $K=p_1+p_2=p_d$, $\ell = (-K + 2 p_3)/3$;
$d$ counts the diquarks participating in the baryon\footnote{In $\Delta^+$-baryons, the sum ranges over isovector-pseudovector $\{uu\}$, $\{ud\}$ correlations; and in $\Delta^0$-baryons, $\{ud\}$ and $\{dd\}$.  Assuming isospin-symmetry, as we do throughout, the correlation amplitudes and propagators are identical for all these diquarks.}
and
$\Gamma^d_{1^+\mu}$, $\Delta^{1^+ d}_{\mu\nu}$ are, respectively, the associated correlation amplitude and propagator;
$u_\rho(P)$ is a Rarita-Schwinger spinor (Ref.\,\cite{Segovia:2014aza}, Appendix\,B);
and
\begin{subequations}
\label{10d}
\begin{align}
\varphi_{\nu\rho}^{\Delta d}(\ell;P) & =
\sum_{k=1}^8 {\mathpzc a}_{\Delta k}^{d}(\ell^2,\ell\cdot P)\, {\mathpzc D}^k_{\nu\rho}(\ell;P)\,,\\
& {\mathpzc D}^k_{\nu\rho} = {\mathpzc S}^k\,\delta_{\nu\rho}\,,   \rule{8ex}{0ex} k=1,2\,,\\
& {\mathpzc D}^k_{\nu\rho} = i\gamma_5\,{\mathpzc A}_\nu^{k-2}\,\ell^\perp_\rho\,, \quad k=3,\dots,8\,,
\end{align}
\end{subequations}
with
\begin{align}
\label{diracbasis}
\nonumber
{\mathpzc S}^1 & = {\mathbf I}_{\rm D} \,,\;
{\mathpzc S}^2  = i \gamma\cdot\hat\ell - \hat\ell \cdot\hat P {\mathbf I}_{\rm D}\,, \\
{\mathpzc A}^1_\nu & =  \gamma\cdot\ell^\perp \hat P_\nu\,,\;
{\mathpzc A}^2_\nu  = - i \hat P_\nu {\mathbf I}_{\rm D}\,,\;
{\mathpzc A}^3_\nu  = \gamma\cdot\hat\ell^\perp \hat\ell^\perp_\nu\,, \\
\nonumber
{\mathpzc A}^4_\nu & = i\hat \ell_\nu^\perp {\mathbf I}_{\rm D}\,,\;
{\mathpzc A}^5_\nu = \gamma_\nu^\perp - {\mathpzc A}^3_\nu\,,\;
{\mathpzc A}^6_\nu = i \gamma_\nu^\perp \gamma\cdot\hat\ell^\perp - {\mathpzc A}^4_\nu\,,
\end{align}
%}
$\hat\ell^2=1$, $\hat P^2 = -1$, $\ell^\perp = \hat\ell_\nu +\hat\ell\cdot\hat P \hat P_\nu$, $\gamma^\perp = \gamma_\nu +\gamma\cdot\hat P \hat P_\nu$.
The (unamputated) Faddeev wave function, $\chi(\ell^2,\ell\cdot P; P^2)$, can be computed from the amplitude specified by Eqs.\,\eqref{FaddeevAmp10}\,--\,\eqref{10d}
%, \eqref{8sa}, \eqref{FaddeevAmp10}, \eqref{10d}
%
simply by attaching the appropriate dressed-quark and diquark propagators.  It may also be decomposed in the form of Eqs.\,\eqref{10d}.  Naturally, the scalar functions are different, and we label them $\tilde {\mathpzc a}_{\Delta k}^{d}$.

\begin{figure*}[!t]
\begin{center}
\begin{tabular}{lr}
\includegraphics[clip,width=0.42\linewidth]{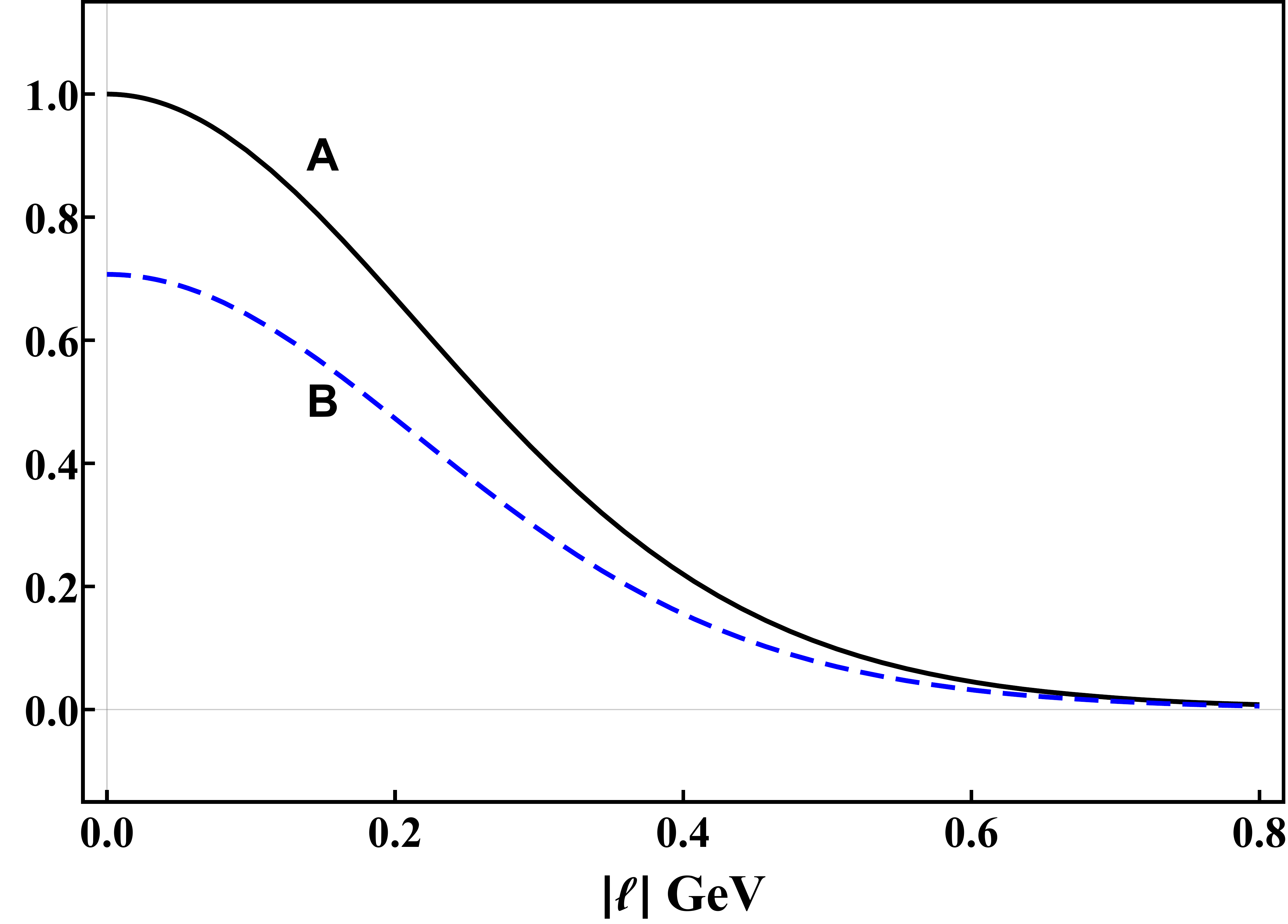}\hspace*{2ex } &
\includegraphics[clip,width=0.42\linewidth]{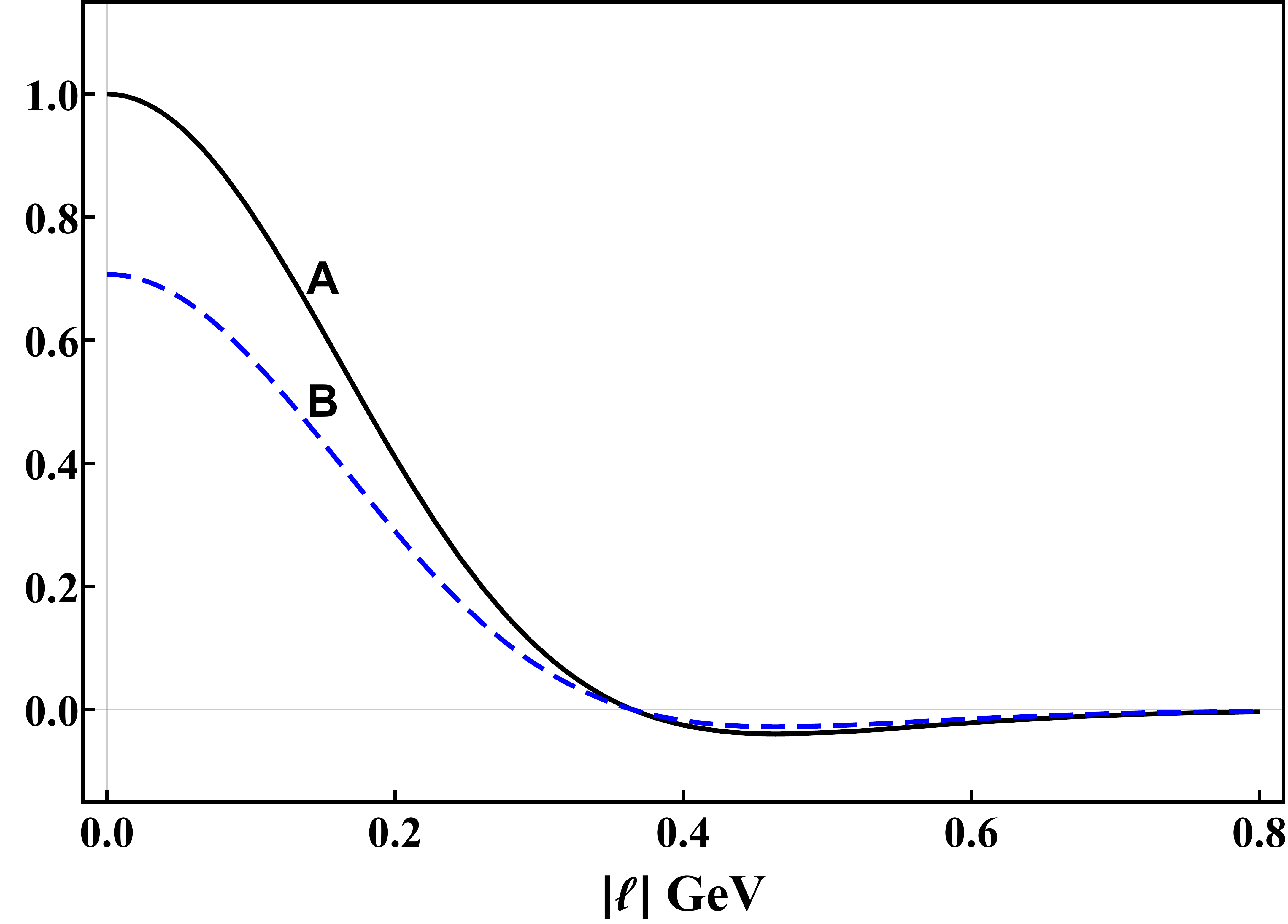}\vspace*{-0ex}
\end{tabular}
\begin{tabular}{lr}
\hspace*{-1.5ex}\includegraphics[clip,width=0.44\linewidth]{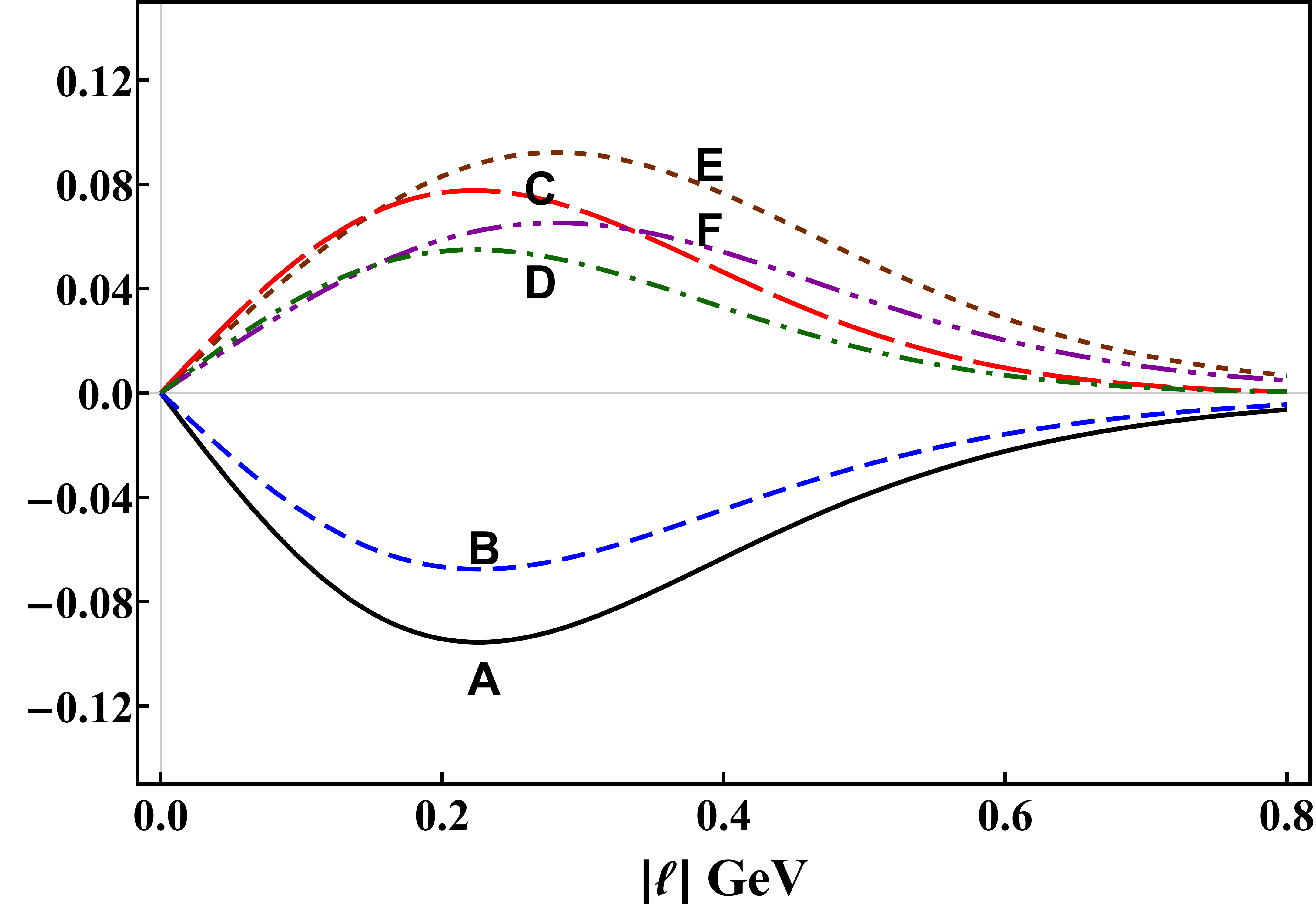}\hspace*{0ex} &
\includegraphics[clip,width=0.44\linewidth]{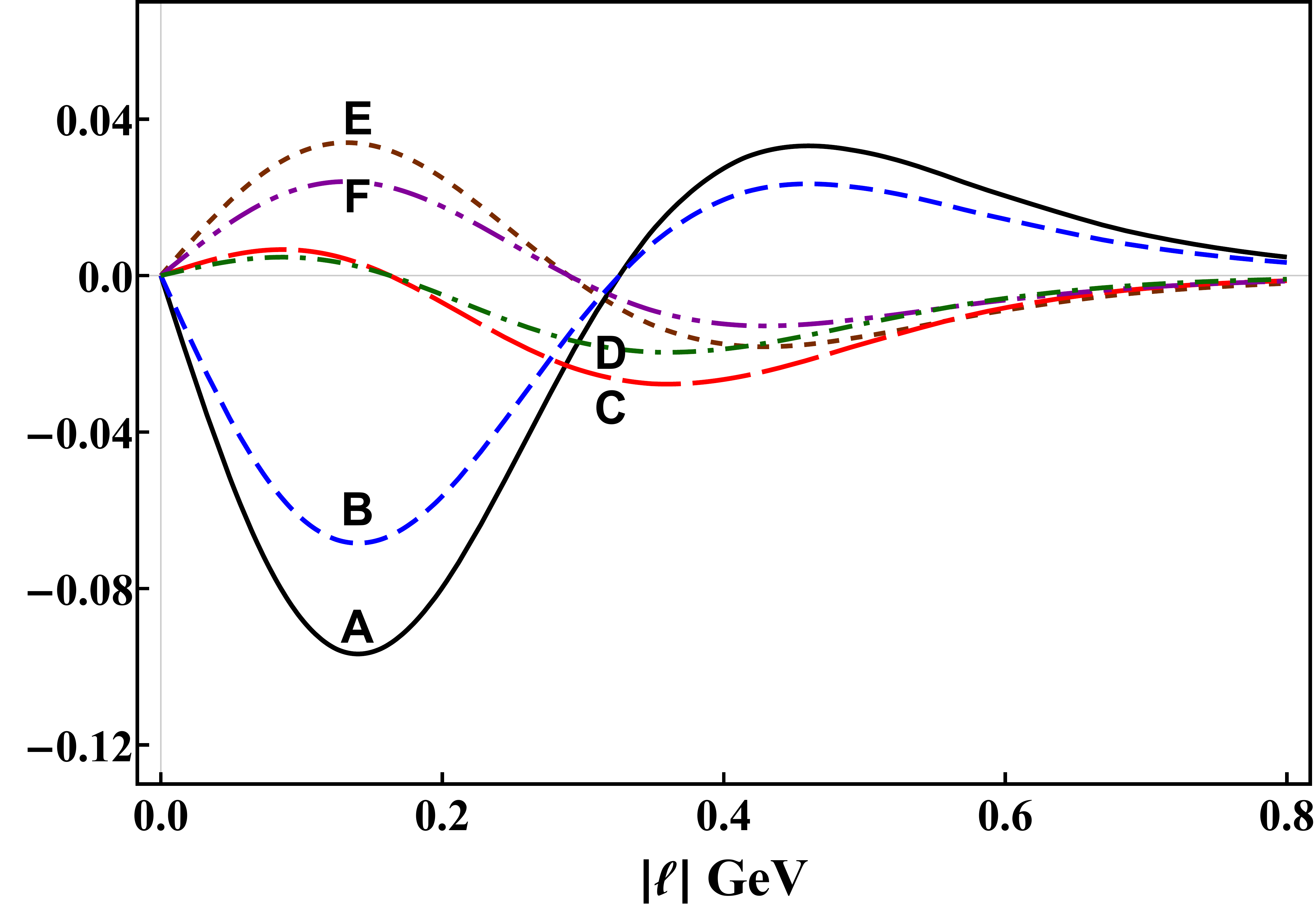}\vspace*{-0ex}
\end{tabular}
\begin{tabular}{lr}
\hspace*{-1ex}\includegraphics[clip,width=0.44\linewidth]{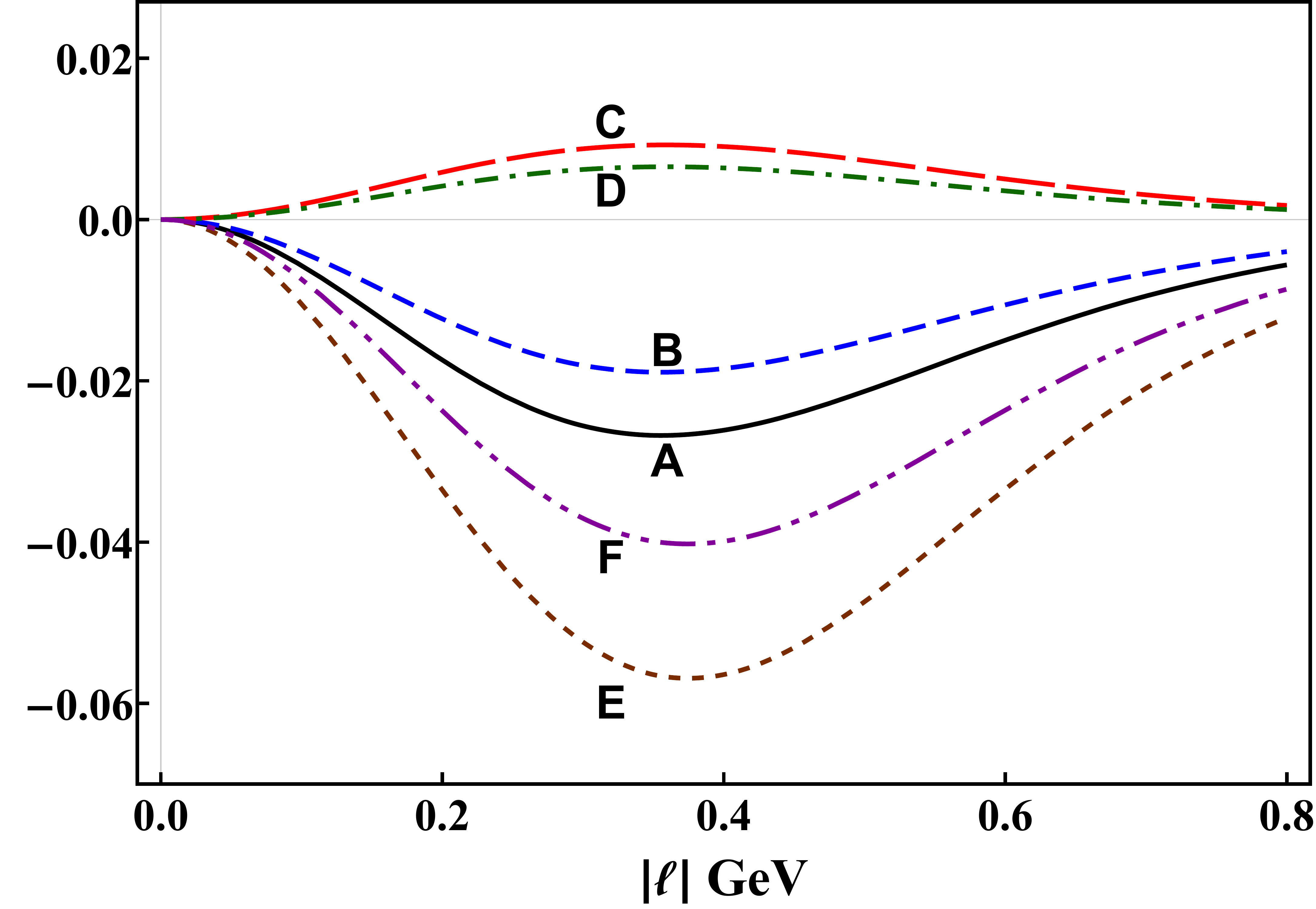}\hspace*{2ex } &
\includegraphics[clip,width=0.42\linewidth]{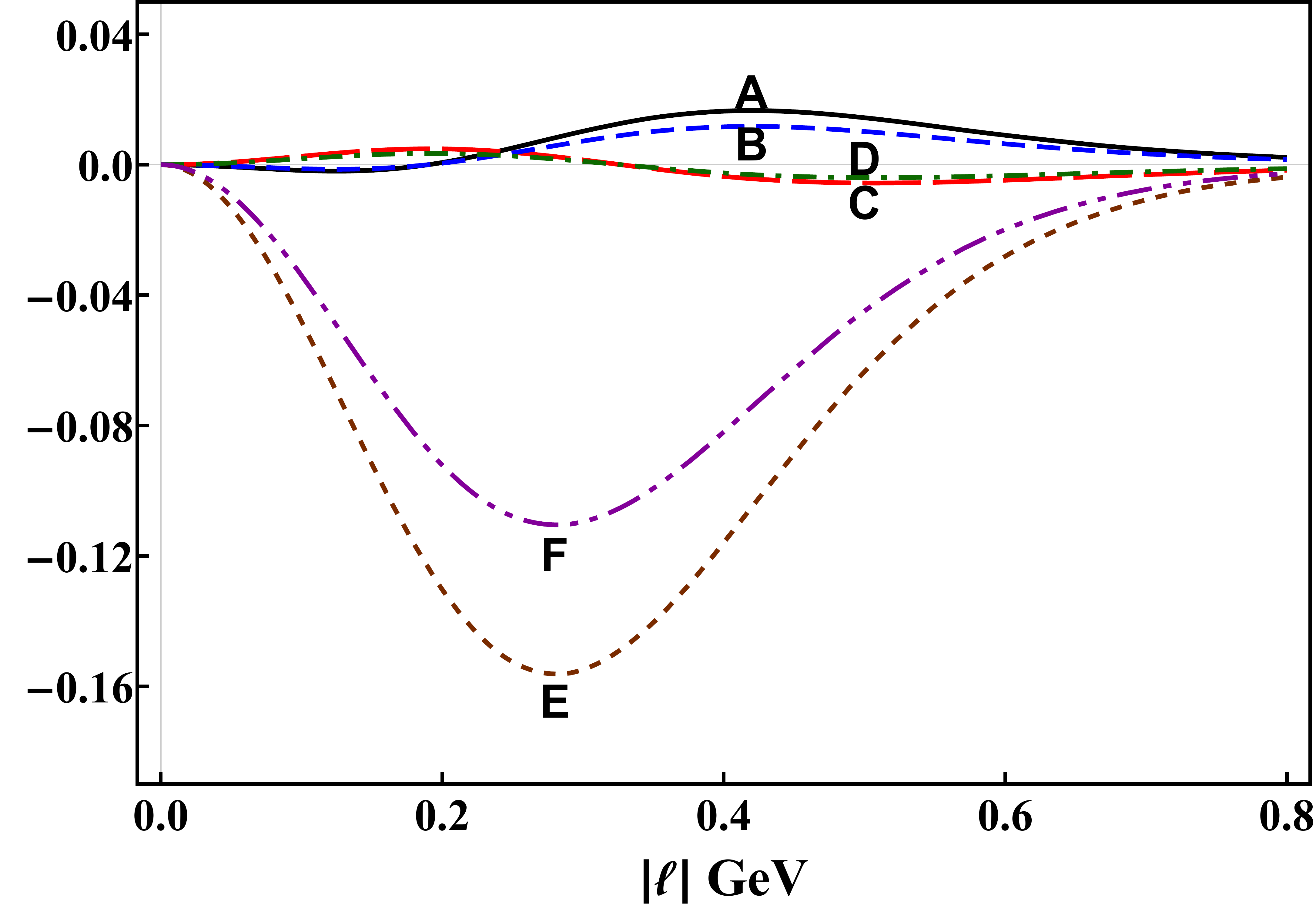}\vspace*{-0ex}
\end{tabular}
\end{center}
\caption{\label{10xistarS}
Faddeev wave functions of $\Delta^+$ baryons, zeroth Chebyshev-moment projections, Eq.\,\eqref{Wproject}:
\emph{left panels} -- ground state; and \emph{right panels} -- first positive-parity excitation.
Superscripts: ``$0$'' labels the pseudovector $(I=1,I_z=0)$ $\{ud\}$-diquark; and ``$1$'' labels the pseudovector $(I=1,I_z=1)$ $\{uu\}$-diquark.  In the isospin-symmetry limit, the associated solution functions satisfy ${\mathpzc a}_{k}^0 = \sqrt{2} {\mathpzc a}_{k}^1$, $k=1,\ldots,8$.
\underline{$S$-wave}: top row. Legend.
``A'' $ \to \tilde{\mathpzc a}_{1}^0+(-\tilde{\mathpzc a}_{6}^0+\tilde{\mathpzc a}_{8}^0)/3$; and
``B'' $\to \tilde{\mathpzc a}_{1}^1+(-\tilde{\mathpzc a}_{6}^1+\tilde{\mathpzc a}_{8}^1)/3$.
\underline{$P$-wave}: middle row. Legend.
``A'' $\to \tilde{\mathpzc a}_{4}^0$;  % F
``B'' $\to \tilde{\mathpzc a}_{4}^1$;  % E
``C'' $\to (2\tilde{\mathpzc a}_{2}^0-\tilde{\mathpzc a}_{5}^0-2\tilde{\mathpzc a}_{7}^0)/3$; % A
``D'' $\to (2\tilde{\mathpzc a}_{2}^1-\tilde{\mathpzc a}_{5}^1-2\tilde{\mathpzc a}_{7}^1)/3$; % C
``E'' $\to \tilde{\mathpzc a}_{2}^0-(\tilde{\mathpzc a}_{5}^0-\tilde{\mathpzc a}_{7}^0)/5$; and % D
``F'' $\to \tilde{\mathpzc a}_{2}^1-(\tilde{\mathpzc a}_{5}^1-\tilde{\mathpzc a}_{7}^1)/5$;  % B
\underline{$D$-wave}: bottom row. Legend.
``A'' $\to\tilde{\mathpzc a}_{3}^0$; %C
``B'' $\to\tilde{\mathpzc a}_{3}^1$; % E
``C'' $\to-(\tilde{\mathpzc a}_{6}^0+2\tilde{\mathpzc a}_{8}^0)/3$; % A
``D'' $\to-(\tilde{\mathpzc a}_{6}^1+2\tilde{\mathpzc a}_{8}^1)/3$;  % B
``E'' $\to-\tilde{\mathpzc a}_{6}^0+\tilde{\mathpzc a}_{8}^0$; and  % F
``F'' $\to-\tilde{\mathpzc a}_{6}^1+\tilde{\mathpzc a}_{8}^1$.          % D
$F$-wave components are negligible for all decuplet baryons \cite{Chen:2019fzn}.
}
\end{figure*}

In order to visualise the wave function solutions of the Faddeev equations, it is convenient to work with univariate Chebyshev projections of the scalar functions used to express them.  Furthermore, as with mesons \cite{Maris:1997tm, Maris:1999nt}, one usually focuses on the zeroth functional moment of the given function ($u=\ell\cdot P/\sqrt{\ell^2 P^2}$):
\begin{equation}
\label{Wproject}
%\textstyle
\tilde{\mathpzc a}(\ell^2;P^2) = \frac{2}{\pi} \int_{-1}^1 \! du\,\sqrt{1-u^2}\,
\tilde{\mathpzc a}(\ell^2,u; P^2)\,,
\end{equation}
because it is typically dominant in realistic solutions and hence expresses the largest amount of information.

The order-zero Chebyshev projections of the Faddeev wave function for the proton are plotted in Ref.\,\cite{Chen:2017pse}, Fig.\,4; and our calculation reproduces those results.
% ... Chen's conventions are the same, but he has distinguished {uu} and {ud} and plotted each separately, leading to an extra curve and a difference in sign that is explained by careful accounting ... EMail Sun. 30/12/18
Herein, therefore, in Fig.\,\ref{10xistarS} we depict the projections for the ground-state $\Delta$-baryon and its first positive-parity excitation.  In all cases, we plot that combination of functions which has a well-defined value of quark-diquark orbital angular momentum in the baryon's rest frame.
A key observation here is that, for the ground-state, each projection is of a unique sign (positive or negative).  On the other hand, with the exception of two $D$-wave components $(-\tilde{\mathpzc a}_{6}^{0,1}+\tilde{\mathpzc a}_{8}^{0,1})$, all excited-state projections possess a single zero.  As noted elsewhere \cite{Segovia:2015hra, Chen:2017pse, Mezrag:2017znp, Qin:2018dqp}, this pattern of behaviour indicates that the positive-parity excitation may be interpreted as the simplest radial excitation of its ground-state partner.

Figure~\ref{10xistarS} also shows that the $\Delta$-baryon ground-state and positive-parity excitation are primarily $S$-wave in character: the magnitudes of the curves in the top row are greater than those in the other rows.  Naturally, we replicate the results of Ref.\,\cite{Chen:2019fzn}, \emph{viz}.\ the ground-state mass is almost insensitive to non-$S$-wave components; and in the first positive-parity excitation, $P$-wave components generate a little repulsion, some attraction is provided by $D$-waves, and $F$-waves have no measurable impact.  Evidently, too, some $S$-wave strength is shifted into $P$- and $D$-wave contributions within the positive-parity excitation \cite{Eichmann:2016nsu, Qin:2018dqp, Chen:2019fzn}.  Notwithstanding their smaller magnitudes, we will see that the higher partial-waves have noticeable effects on electroproduction form factors.

Let us return to the masses in Eq.\,\eqref{eqMasses}.  Empirical values of these pole locations are \cite{Tanabashi:2018oca} (in GeV): $0.939$, $1.21 - i 0.05$, $1.51 - i 0.14$.
(The physical $\Delta$-baryons are unstable and hence the associated pole has an imaginary part.)
At first glance, these values appear unrelated to those in Eq.\,\eqref{eqMasses}.  However, deeper consideration reveals \cite{Eichmann:2008ae, Eichmann:2008ef} that the kernel in Fig.\,\ref{figFaddeev} has an intrinsic weakness: resonant contributions , \emph{viz}.\ meson-baryon final-state-interactions (MB\,FSIs), are omitted.  It is such effects which are resummed in dynamical coupled channels models, generating the widths and thereby transforming bare-baryons into the observed states \cite{Kamano:2018sfb, Doring:2018kue}.
Our Faddeev equation should therefore be understood as producing the dressed-quark core of the bound-state, not the completely-dressed object.

Clothing the nucleon's dressed-quark core by including resonant contributions to the kernel produces a physical nucleon whose mass is $\approx 0.2$\,GeV lower than that of the core \cite{Ishii:1998tw, Hecht:2002ej}.
%; and $1.19-0.2 = 0.99 \approx 0.94\,$GeV.
Similarly, MB\,FSIs reduce the $\Delta(1232)$-baryon's core mass by $\approx 0.16\,$GeV \cite{JuliaDiaz:2007kz, JuliaDiaz:2007fa, Suzuki:2009nj} and the Roper resonance's core-mass by $0.3\,$GeV \cite{Suzuki:2009nj}.
Evidently, such reductions shift the mass of a given baryon's dressed-quark core into alignment with the measured Breit-Wigner mass of the associated physical states.  Moreover, this pattern is seen to prevail broadly, extending to baryons in the multiplets of flavour-$SU(5)$  \cite{Qin:2019hgk, Yin:2019bxe}.

Our approach thus delivers the dressed-quark-core contribution to a given observable and that this should subsequently be corrected by incorporating MB\,FSIs.  These features have long been appreciated and exploited in developing a successful body of work on the baryon spectrum and elastic and transition form factors, \emph{e.g}.\ Refs.\,\cite{Segovia:2014aza, Chen:2019fzn}; and we capitalise on such experience herein.

%Resonance electroproduction

%\cite{Aznauryan:2015zta, Aznauryan:2016wwm}  second reference indicates that configuration mixing has a big impact.

%$\Delta(1600)$ has large $D$-wave component in rest-frame wave function \ldots we show it affects the transition form factors \ldots but it is nevertheless an $S$-wave state because there is no bound-state unless $S$-wave piece is present.

\section{Transition Current}
\label{SecTransition}
Electromagnetic \mbox{$N\to\Delta$} transitions are described by three form factors \cite{Jones:1972ky}: magnetic-dipole, $G_M^\ast$; electric quadrupole, $G_E^\ast$; and Coulomb (longitudinal) quadrupole, $G_C^\ast$.  They arise through consideration of the transition current:
\begin{equation}
J_{\mu\lambda}(K,Q) =
\Lambda_{+}(P_{f})R_{\lambda\alpha}(P_{f})i\gamma_{5}\Gamma_{\alpha\mu}(K,Q)\Lambda_
{+}(P_{i}),
\label{eq:JTransition}
\end{equation}
where: $P_{i}$, $P_{f}$ are, respectively, the incoming nucleon and outgoing $\Delta$ momenta, $P_{i}^{2}=-m_{N}^{2}$, $P_{f}^{2}=-m_{\Delta}^{2}$; $Q_\mu=(P_{f}-P_{i})_\mu$ is the incoming photon momentum, $K=(P_{i}+P_{f})/2$; and $\Lambda_{+}(P_{i})$, $\Lambda_{+}(P_{f})$ are, respectively, positive-energy projection operators for the nucleon and $\Delta$, with the Rarita-Schwinger tensor projector $R_{\lambda\alpha}(P_f)$ arising in the latter connection.  (See Ref.\,\cite{Segovia:2014aza}, Appendix\,B.)

In order to succinctly express $\Gamma_{\alpha\mu}(K,Q)$, we define
\begin{equation}
\check K_{\mu}^{\perp} = {\cal T}_{\mu\nu}^{Q} \check{K}_{\nu}
= (\delta_{\mu\nu} - \check{Q}_{\mu} \check{Q}_{\nu}) \check{K}_{\nu},
\end{equation}
with $\check{K}^{2} = 1 = \check{Q}^{2}$, in which case
{\allowdisplaybreaks
\begin{align}
\nonumber
\Gamma_{\alpha\mu}(K,Q) =
\mathpzc{k}
\left[\frac{\lambda_m}{2\lambda_{+}}(G_{M}^{\ast}-G_{E}^{\ast})\gamma_{5}
\varepsilon_{\alpha\mu\gamma\delta} \check K_{\gamma}\check{Q}_{\delta}\right.  \\
\left. -
G_{E}^{\ast} {\cal T}_{\alpha\gamma}^{Q} {\cal T}_{\gamma\mu}^{K}
- \frac{i\varsigma}{\lambda_m}G_{C}^{\ast}\check{Q}_{\alpha} \check
K^\perp_{\mu}\right],
\label{eq:Gamma2Transition}
\end{align}}
\hspace*{-0.5\parindent}where
$\mathpzc{k} = \sqrt{(3/2)}(1+m_\Delta/m_N)$,
$\varsigma = Q^{2}/[2\Sigma_{\Delta N}]$,
$\lambda_\pm = \varsigma + t_\pm/[2 \Sigma_{\Delta N}]$
with $t_\pm = (m_\Delta \pm m_N)^2$,
$\lambda_m = \sqrt{\lambda_+ \lambda_-}$,
$\Sigma_{\Delta N} = m_\Delta^2 + m_N^2$, $\Delta_{\Delta N} = m_\Delta^2 - m_N^2$.

With a concrete expression for the current in hand, one may obtain the form factors using any three sensibly chosen projection operations, \emph{e.g}.\ with \cite{Eichmann:2011aa}
\begin{subequations}
\label{ProjectionsE}
\begin{align}
\mathpzc{t}_{1} & = \mathpzc{n}
\frac{\sqrt{\varsigma(1+2\mathpzc{d})}}{\mathpzc{d}-\varsigma}
{\cal T}^{K}_{\mu\nu}\check K^\perp_{\lambda} {\rm tr}
\gamma_{5}J_{\mu\lambda}\gamma_{\nu}\,, \\
\mathpzc{t}_{2} & = \mathpzc{n} \frac{\lambda_{+}}{\lambda_m} {\cal T}^{K}_{\mu\lambda}
{\rm tr} \gamma_{5} J_{\mu \lambda}\,,\\
\mathpzc{t}_{3} & =  3 \mathpzc{n}
\frac{\lambda_+}{\lambda_m}\frac{(1+2\mathpzc{d})}{\mathpzc{d}-\varsigma} \check
K^\perp_{\mu}\check K^\perp_{\lambda} {\rm tr}\gamma_{5}J_{\mu\lambda} \,,
\end{align}
\end{subequations}
where $\mathpzc{d}=\Delta_{\Delta N}/[2 \Sigma_{\Delta N}]$,
$\mathpzc{n}= \sqrt{1-4\mathpzc{d}^{2}}/[4i\mathpzc{k}\lambda_m]$), then
\begin{equation}
\label{GMGEGC}
G_{M}^{\ast} = 3
\left[ \mathpzc{t}_{2}+\mathpzc{t}_{1}\right]\,, \;
G_{E}^{\ast} = \mathpzc{t}_{2}-\mathpzc{t}_{1}\,, \;
G_{C}^{\ast} = \mathpzc{t}_{3}.
\end{equation}
%Like the nucleon's Dirac and Pauli form factors, t123 are probably positive-definite, but that can't be guaranteed.  F1* F2* for N->Roper are not positive definite.
%% t3 <-> G0
%% t1 <-> G+
%% t2 <-> - Sqrt[3] G-
%% t2 ~ G- is suppressed at large Q^2.

The following ratios are often considered in connection with $\gamma^\ast N\to \Delta$ transitions:
\begin{equation}
\label{eqREMSM}
R_{\rm EM} = -\frac{G_E^{\ast}}{G_M^{\ast}}, \quad
R_{\rm SM} = - \frac{|\vec{Q}|}{2 m_\Delta} \frac{G_C^{\ast}}{G_M^{\ast}}
=  - \frac{\lambda_m}{m_\Delta} \frac{G_C^{\ast}}{G_M^{\ast}}\,.
\end{equation}
Since they are identically zero in $SU(6)$-symmetric constituent-quark models, they can be read as measures of deformation in one or both of the hadrons involved.

Following Refs.\,\cite{Oettel:1999gc, Segovia:2014aza}, the transition current in Eq.\,\eqref{eq:JTransition} can be explicated as follows:
\begin{align}
\nonumber
 J_{\mu,\alpha}(P_f,& P_i) =  \sum_{n=1}^6
\int\frac{d^4p}{(2\pi)^4} \frac{d^4k}{(2\pi)^4} \; \\
& \times
\bar\Psi_\alpha(-p;P_f) \, J_{\mu}^n(p,P_f,k,P_i) \, \Psi(k;P_i)\,,
\label{JDeltaNExplicit}
\end{align}
where $\Psi_\alpha$, $\Psi$ are, respectively, the $\Delta$ and nucleon Faddeev amplitudes described in Sec.\,\ref{SecFaddeev}; and the sum ranges over the six diagrams depicted and detailed in Ref.\,\cite{Segovia:2014aza}, Appendix~C.  Each term in  Eq.\,\eqref{JDeltaNExplicit} can be evaluated using standard algebraic and numerical techniques, and results for the form factors obtained subsequently via the projections in Eqs.\,\eqref{ProjectionsE} and combinations in Eqs.\,\eqref{GMGEGC}.

\begin{table}[t]
\caption{\label{static}
Static properties computed from the $\Delta^{+}(1232)$ and $\Delta^{+}(1600)$ elastic form factors.  An empirical value of $G_{M1}(0)$ is available for the $\Delta^{+}(1232)$ \cite{Tanabashi:2018oca}:
$3.6^{+1.3}_{-1.7}\pm 2.0 \pm 4$.  Point-particle values for $J=3/2$ states are: $G_{M1}(0)=3$, $G_{E2}(0)=-3$, $G_{M3}(0)=-1$. \cite{Lorce:2009bs}.
All radii listed in units of the quark-core proton charge radius, $r_p= 0.61\,$fm.
}
\begin{center}
\begin{tabular*}%{|c|c|c|c|c|c|c|}\hline
{\hsize}
{
l@{\extracolsep{0ptplus1fil}}|
c@{\extracolsep{0ptplus1fil}}
c@{\extracolsep{0ptplus1fil}}
c@{\extracolsep{0ptplus1fil}}
c@{\extracolsep{0ptplus1fil}}
c@{\extracolsep{0ptplus1fil}}
c@{\extracolsep{0ptplus1fil}}
c@{\extracolsep{0ptplus1fil}}}\hline\hline
baryon & $r_E\ $ & $G_{M1}(0)\ $ & $r_{M1}\ $ & $G_{E2}(0)\ $ & $r_{E2}\ $ & $G_{M3}(0)\ $ & $r_{M3}\ $ \\\hline
$\Delta^{+}(1232)\ $ &$1.23\ $ & $2.86\ $ & $1.10\ $ & $-6.67\ $ & $1.20\ $ & $-3.00\ $ & $0.48\ $ \\
$\Delta^{+}(1600)\ $ &$1.68\ $ & $1.50\ $ &$1.05\ $ & $-3.00\ $ & $0.79\ $ & $\phantom{-}0.80\ $& $0.64\ $ \\
\hline\hline
\end{tabular*}
\end{center}
\end{table}

In these calculations, the proton and $\Delta^+$-baryon Faddeev amplitudes must be canonically normalised. This is achieved by computing the elastic electric form factor in each case and rescaling the amplitude such that the associated $Q^2=0$ value (electric charge) is unity \cite{Segovia:2014aza}.  Given this necessity, we computed the low-$Q^2$ behaviour of all elastic form factors for each baryon and report the associated static properties of their dressed-quark cores in Table~\ref{static}.  These results lead to the following observations:
\begin{equation}
r_E^{\Delta(1600)} \approx 1.4\, r_E^{\Delta(1232)} ,  \;
r_{M1}^{\Delta(1600)} \approx 0.95\, r_{M1}^{\Delta(1232)} ,
\end{equation}
which may sensibly be compared with $r_E^{\rm Roper} \approx 1.8\, r_p$, $r_M^{\rm Roper} \approx 1.6\, r_M^p$ \cite{Segovia:2015hra};
and the octupole moments of the $\Delta(1232)$ and $\Delta(1600)$ have opposite signs, an outcome that signals the impact of differences in the distribution and strength of higher partial-waves in the respective wave functions (see Fig.\,\ref{10xistarS} herein and the discussion of Fig.\,7 in Ref.\,\cite{Segovia:2014aza}).
In addition, we find that the $\Delta(1600)$ elastic electric form factor possesses a zero, at $Q^2\approx 1.8 \, m_p^2$.  For the $\Delta(1232)$, this zero lies at $Q^2\approx 2.7 \, m_p^2$ \cite{Segovia:2014aza}.  Notably, the ordering and locations are consistent with the electric radii reported in Table~\ref{static}.

\section{Calculated Form Factors: $\mathbf{\Delta(1232)}$}
\label{SecTFFs1232}
\begin{figure}[t]
\centerline{%
\includegraphics[clip, width=0.42\textwidth]{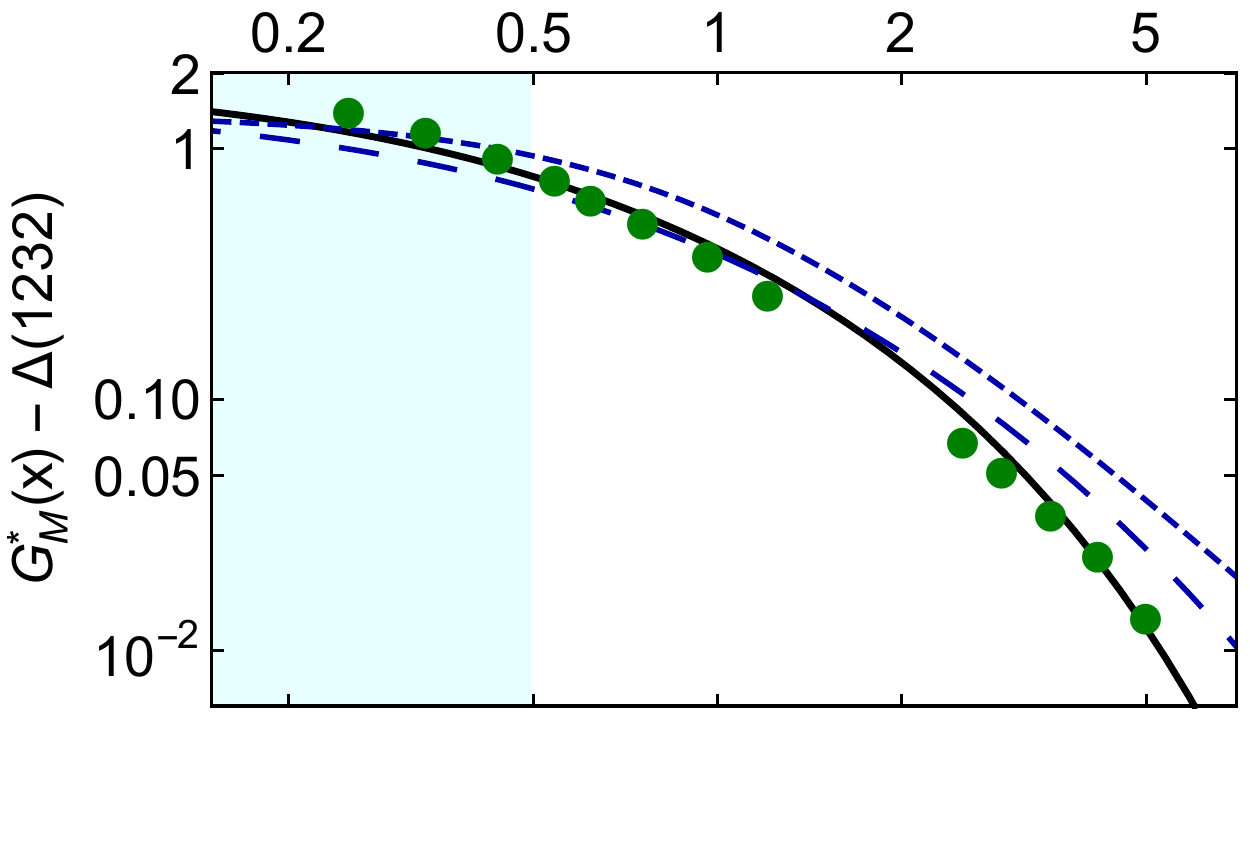}}
\vspace*{-8ex}

\centerline{%
\includegraphics[clip, width=0.42\textwidth]{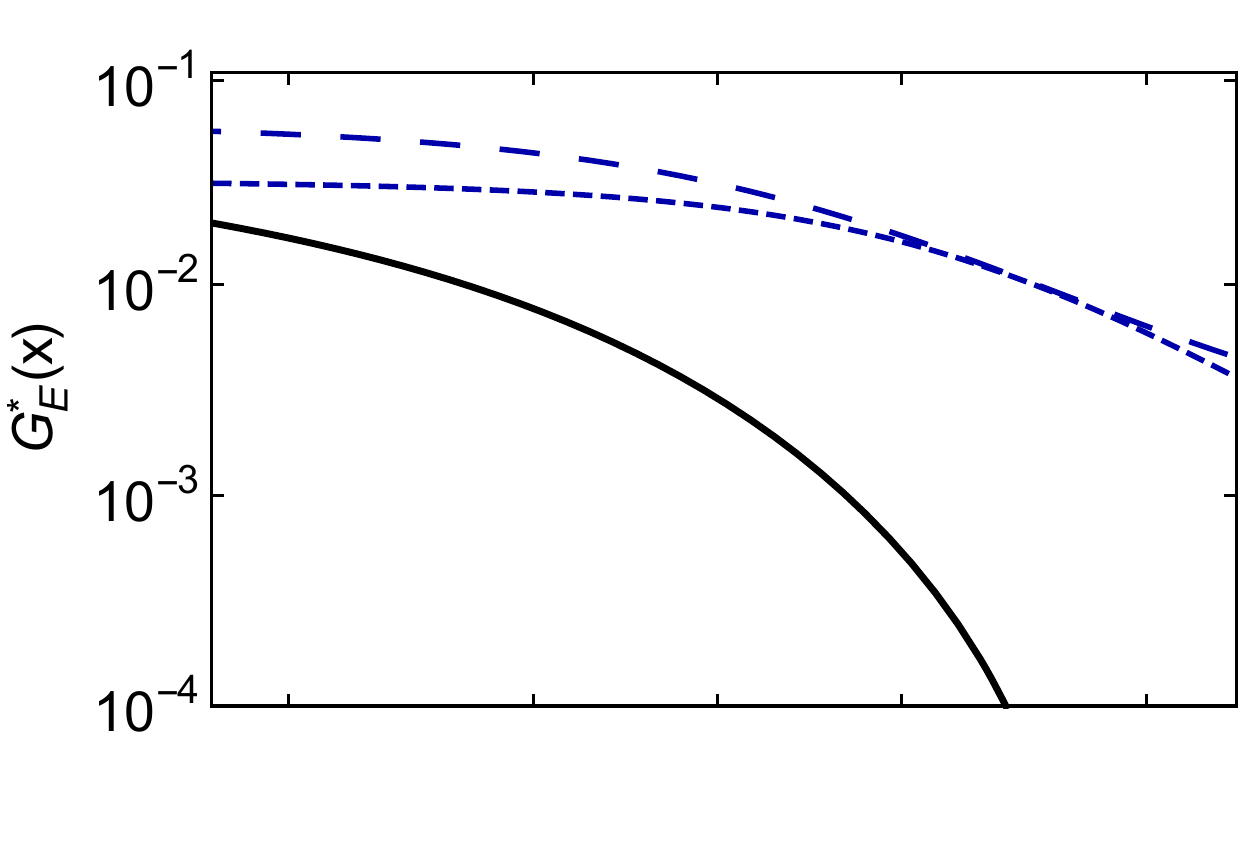}}
\vspace*{-8ex}

\centerline{%
\hspace*{-0.5ex}\includegraphics[clip, width=0.42\textwidth]{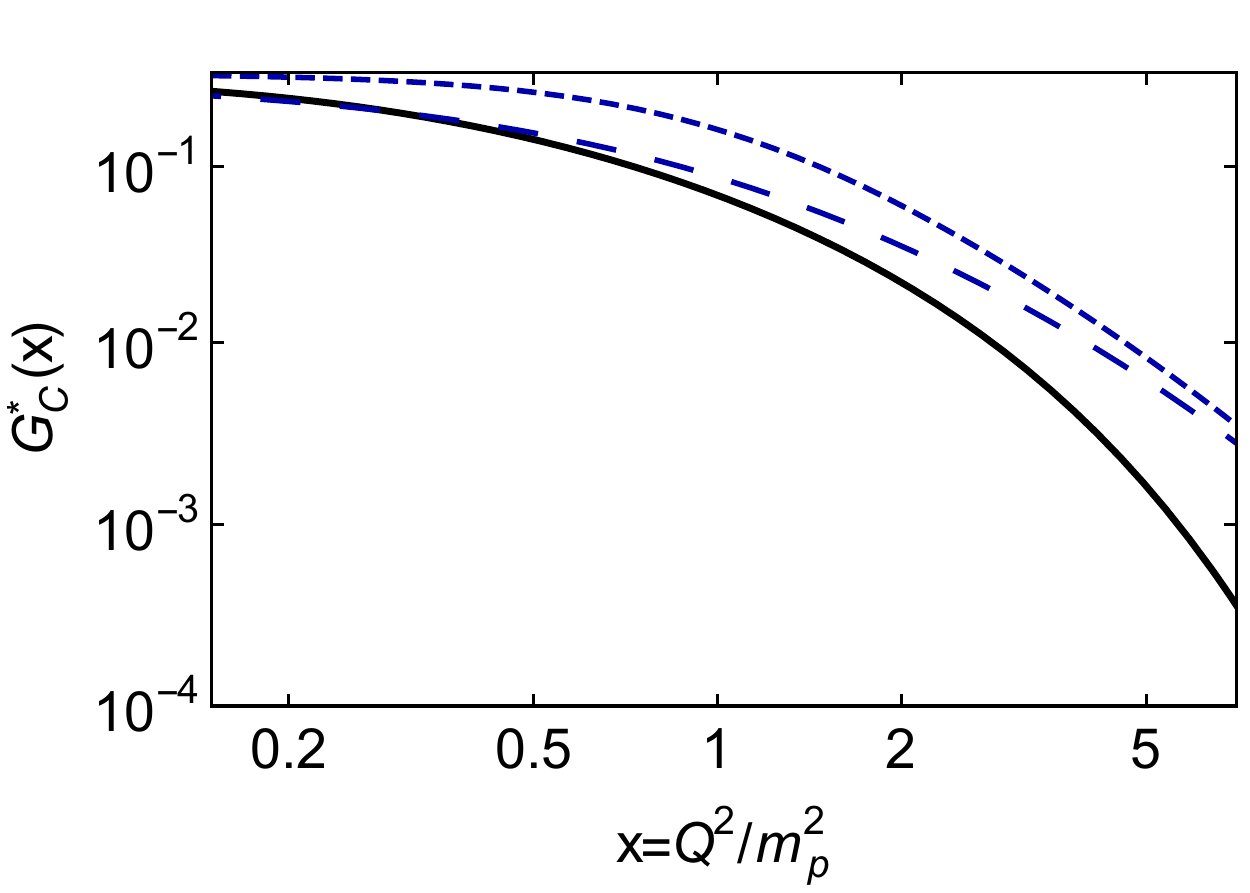}}
\caption{\label{figD1232}
\emph{Top panel}: Magnetic dipole $\gamma^\ast p\to \Delta^+(1232)$ transition form factor compared with contemporary data \cite{Aznauryan:2009mx}.  The conventions of Ref.\,\cite{Jones:1972ky} are employed.
\emph{Middle panel}:  Electric quadrupole transition form factor.
\emph{Bottom panel}:  Coulomb quadrupole transition form factor.
In all panels:
solid (black) curve, complete result;
long-dashed (blue) curve, result obtained when only those components of the $\Delta(1232)$ wave function are retained which correspond to $S$-waves in the rest frame;
and dashed (blue) curve, obtained when both the proton and $\Delta(1232)$ are reduced to $S$-wave states.
}
\end{figure}

Our computed $\gamma^\ast p\to \Delta^+(1232)$ transition form factors are depicted in Fig.\,\ref{figD1232}.  They are accurately interpolated using a simple functional form \cite{Sato:2000jf}:
\begin{equation}
\label{interpolationTFFs}
G_{\mathpzc F}^\ast(x) = \frac{a_0^{\mathpzc F}+a_1^{\mathpzc F} x}{1+ b_1^{\mathpzc F} x + b_2^{\mathpzc F} x^2} {\rm e}^{-c_1^{\mathpzc F} x},
\end{equation}
%GMJS[x_] := (a0 + a1*x)/(1 + b1*x + b2*x^2)*Exp[-c1*x]
with the coefficients given in Table~\ref{TFFinterpolation}.  (These forms should not be used for large-$x$ extrapolation.)

\begin{table}[t]
\caption{\label{TFFinterpolation}
Interpolation coefficients for each of our computed $\gamma^\ast p \to \Delta$ transition form factors, Eq.\,\eqref{interpolationTFFs}.  Blank entries indicate ``0''.
}
\begin{center}
\begin{tabular*}%{|c|c|c|c|c|c|c|}\hline
{\hsize}
{
l@{\extracolsep{0ptplus1fil}}|
l@{\extracolsep{0ptplus1fil}}
c@{\extracolsep{0ptplus1fil}}
c@{\extracolsep{0ptplus1fil}}
c@{\extracolsep{0ptplus1fil}}
c@{\extracolsep{0ptplus1fil}}
c@{\extracolsep{0ptplus1fil}}
c@{\extracolsep{0ptplus1fil}}}\hline\hline
& ${\mathpzc F}\ $ & $a_0^{\mathpzc F}\ $& $a_1^{\mathpzc F}\ $& $b_1^{\mathpzc F}\ $& $b_2^{\mathpzc F}\ $& $c_0^{\mathpzc F}\ $\\\hline
$\Delta(1232)\ $ & $M$ & $\phantom{-}1.93\phantom{01}\ $& $4.15\ $ & $3.92\ $ & $3.85\ $&  $0.55\ $\\
                         & $E$ & $\phantom{-}0.041\phantom{0}\ $& $-0.010\ $ & $4.62\ $ &$0.68\ $ &  $0.55\ $\\
& $C$ & $\phantom{-}0.30\phantom{01}\ $ & $\phantom{-}0.030\ $ &$1.58\ $ & $0.35\ $ & $0.55\ $ \\\hline
$\Delta(1600)\ $ & $M$ & $\phantom{-}0.32\phantom{01}\ $ & & & $0.22\ $& $0.08\ $\\
& $E$ & $-0.022\phantom{1}\ $& & $-0.10\ $& $0.15\ $& $0.45\ $\\
& $C$ & $\phantom{-}0.14\phantom{01}\ $ & & & $0.23\ $& $0.07\ $\\\hline\hline
\end{tabular*}
\end{center}
\end{table}
%% FitStandard.nb
%%b1 -> 4.62446, b2 -> 0.678836

Considering Fig.\,\ref{figD1232}, it is evident both that $G_M^\ast$, the magnetic dipole form factor, dominates this transition and our result agrees with modern data on $Q^2 \gtrsim 0.5\,m_p^2$.  As explained elsewhere \cite{Sato:2000jf, JuliaDiaz:2007kz}, incorporation of MB\,FSIs is crucial to ensuring agreement on $Q^2 \lesssim 0.5\,m_p^2$, \emph{e.g}.\ such effects increase the result by a factor of $\approx 1.5$ at $Q^2=0$.  This ``meson-cloud domain'' is indicated by shading in the top panel of Fig.\,\ref{figD1232}.  Its size typically depends on the baryon(s) being considered, \emph{e.g}. extending to $Q^2\approx 2 m_p^2$ for nucleon elastic form factors \cite{Eichmann:2008ef, Eichmann:2011vu} and Roper electroproduction \cite{Burkert:2017djo}.

The $\gamma^\ast p\to \Delta^+(1232)$ electric and Coulomb quadrupole form factors are small but nonzero, highlighting that the dressed-quark cores of the baryons involved are deformed, \emph{viz}.\ not purely $S$-wave in their rest frames.  Although this is obvious from inspection of their Poincar\'e-covariant wave functions (Ref.\,\cite{Chen:2017pse}, Fig.\,4, and Fig.\,\ref{10xistarS} above), $G_{E,C}^\ast$ are a measurable manifestation of the distortion's magnitude.  (As will subsequently become apparent, it is deformation of the $\Delta$-baryons which is most important.)

Each panel in Fig.\,\ref{figD1232} contains three curves:
the solid (black) curve is our complete prediction;
the long-dashed (blue) curve is obtained when only those components of the $\Delta(1232)$ wave function are retained which correspond to $S$-waves in the rest frame;
and the dashed (blue) curve is obtained when both the proton and $\Delta(1232)$ are reduced to $S$-wave states.
Notably: the role played by higher partial waves in the wave functions increases with momentum transfer (something also observed in meson form factors \cite{Maris:1998hc}), here generating destructive interference;
agreement with data on $G_M^\ast$ is impossible without the higher partial waves;
and the effect of these components is very large in $G_E^\ast$, unsurprisingly, because it is a difference of two positive-definite functions.  (The complete result for $G_E^\ast$ exhibits a zero at $x\approx 4$, which is absent in the S-wave-only result(s).)

\begin{figure}[t]
\centerline{%
\includegraphics[clip, width=0.42\textwidth]{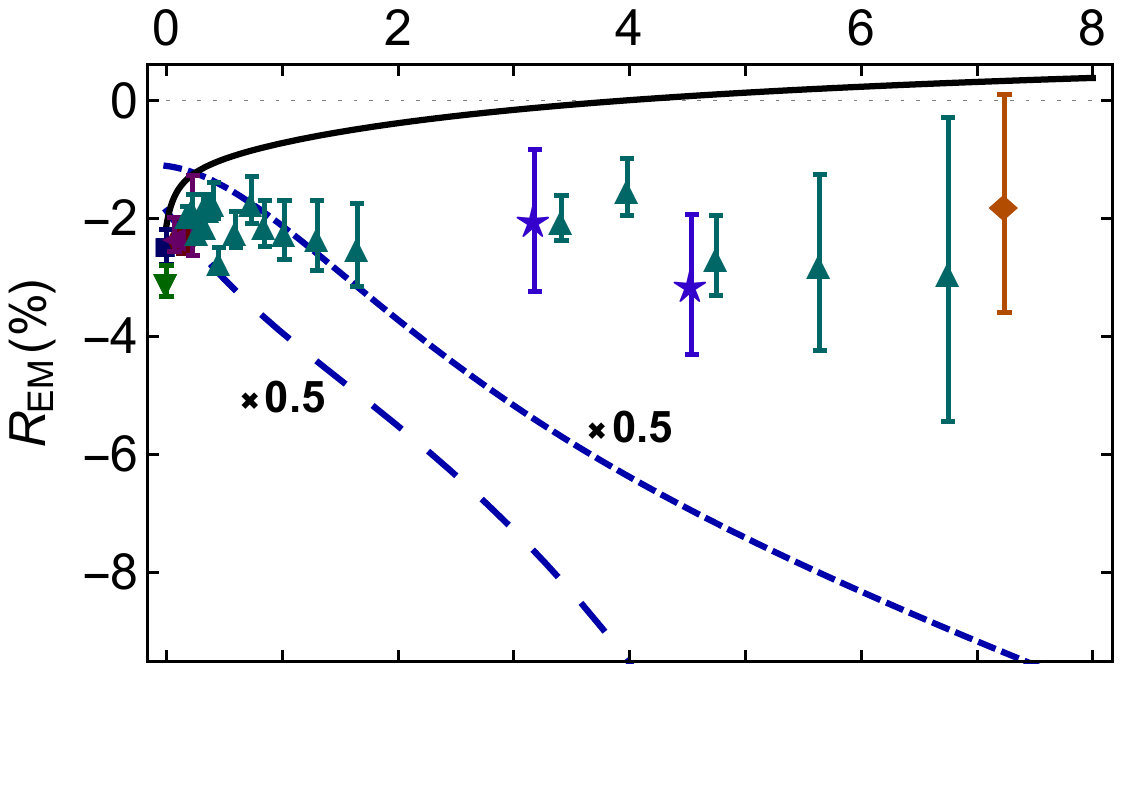}}
\vspace*{-8ex}

\centerline{%
\hspace*{-1.5ex}\includegraphics[clip, width=0.435\textwidth]{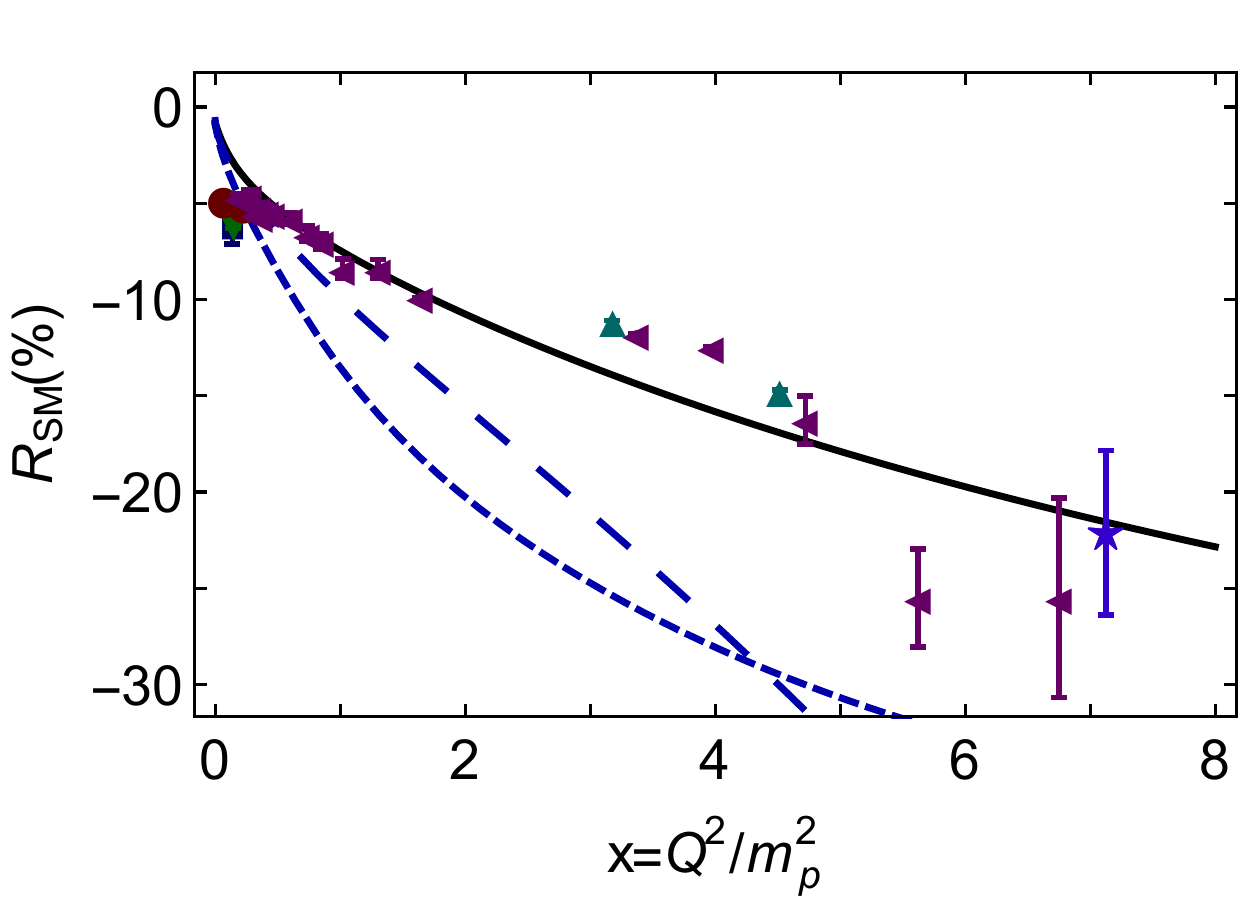}}
\caption{\label{figD1232ratios}
$\gamma^\ast p\to \Delta^+(1232)$ transition, quadrupole ratios in Eq.\,\eqref{eqREMSM}:
\emph{upper panel}, $R_{EM}$ and; \emph{lower panel}, $R_{SM}$.
In all panels:
solid (black) curve, complete result;
long-dashed (blue) curve, result obtained when only those components of the $\Delta(1232)$ wave function are retained which correspond to $S$-waves in the rest frame;
and dashed (blue) curve, obtained when both the proton and $\Delta(1232)$ are reduced to $S$-wave states.
The data in both panels are drawn from Refs.\,\protect\cite{
Beck:1999ge,
Pospischil:2000ad,
Blanpied:2001ae,
Sparveris:2004jn,
Stave:2008aa,
Aznauryan:2009mx}.
}
\end{figure}

%Plot[RSM\[CapitalDelta]1232[xN], {xN, 0, xmx},PlotStyle -> {{Thick, Black}}],
%Plot[RSM\[CapitalDelta]1232S[x], {x, 0, xmx}, PlotStyle -> {Thick, Dashing[Large], Darker[Blue]}],
%Plot[RSM\[CapitalDelta]1232SS[x], {x, 0, xmx},PlotStyle -> {Thick, Dashed, Darker[Blue]}],

In Fig.\,\ref{figD1232ratios}, to further elucidate the observable impacts of higher partial-waves in the Poincar\'e-covariant wave functions,  we depict the ratios $R_{EM}$, $R_{SM}$ defined in Eq.\,\eqref{eqREMSM}.  The long-dashed and dashed curves in the upper panel are each multiplied by $0.5$ so that they fit comfortably within the frame.  The need for such multiplication highlights the substantial impact of higher partial-waves on $R_{EM}$.  Such marked sensitivity of $R_{EM}$ has been observed elsewhere \cite{Eichmann:2011aa, Segovia:2014aza, Segovia:2016zyc}; but the difference between our prediction for the response and that in Ref.\,\cite{Eichmann:2011aa} shows $R_{EM}$ to be particularly susceptible to model-details.
%%%SS ->S adding P-wave in proton has a modest effect
%%... DS*pS
%%... + DS*pP noticeable constructive interference in magnitude
%%... + (DP+DD)*pS + (DP+DD)*pP ... enormous desctructive interference

It is here worth reiterating a conclusion from Ref.\,\cite{Segovia:2016zyc}, \emph{viz}.\ in the $\gamma^\ast p \to\Delta^+(1232)$ transition, $G_E^\ast$ is dominated by terms involving a scalar diquark in the proton and a pseudovector diquark in the $\Delta^+(1232)$, with photon-diquark interactions controlling the transition away from $x=0$.  It follows that, within the dressed-quark core, the electric quadrupole transition proceeds primarily by a photon transforming the $0^+$-diquark in the proton into a $1^+$-diquark ($\delta J=1$) in the $\Delta^+(1232)$, with the overlap of quark-diquark components in the rest-frame Faddeev wave functions of the proton and $\Delta^+(1232)$ that differ by one unit of angular momentum.  This explains why the shift induced by adding $P$- and $D$-waves in the $\Delta(1232)$ is especially large.

Given, too, that axial-vector diquark contributions interfere constructively with MB\,FSIs \cite{Ishii:1998tw, Hecht:2002ej}, then these features also indicate that $G_E^\ast$ should be most sensitive to meson cloud contributions \cite{Segovia:2014aza}.

\begin{figure}[t]
\centerline{%
\includegraphics[clip, width=0.42\textwidth]{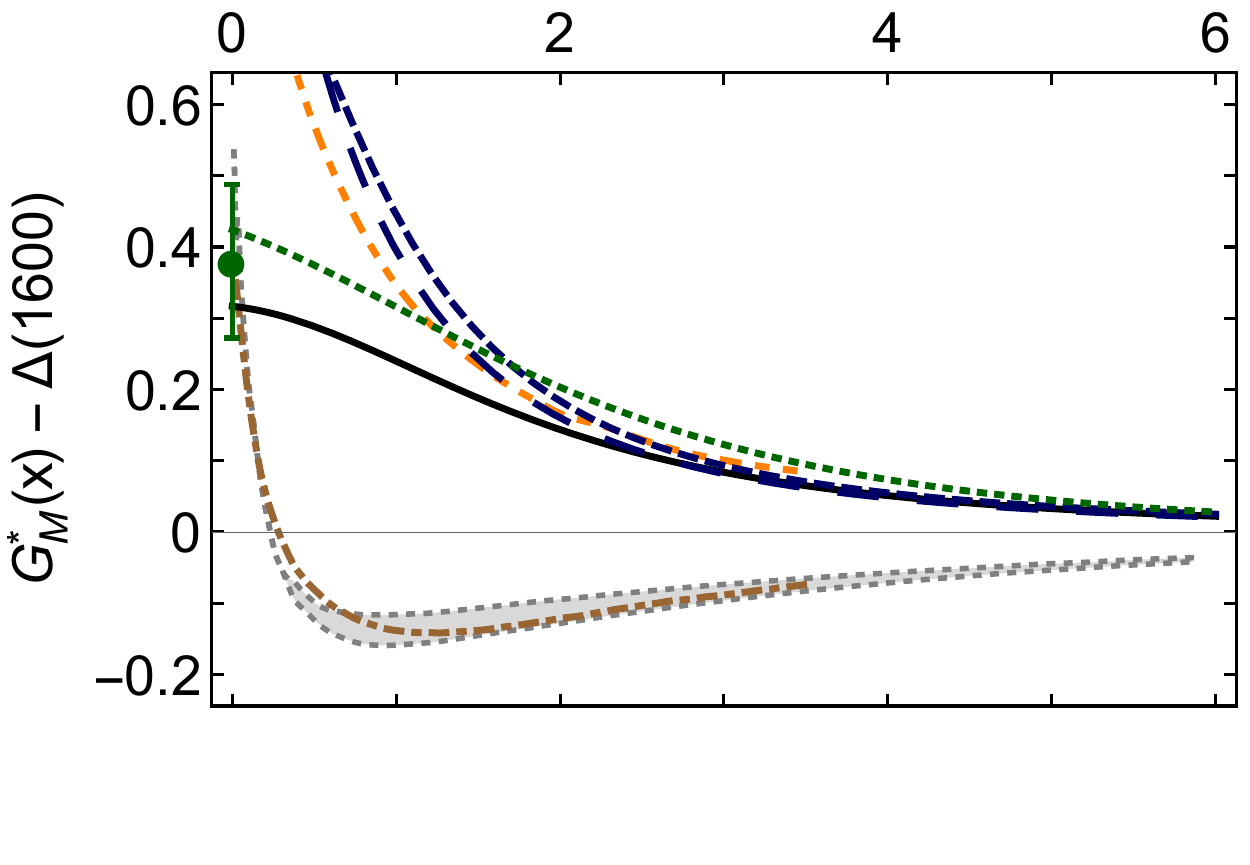}}
\vspace*{-8ex}

\centerline{%
\includegraphics[clip, width=0.42\textwidth]{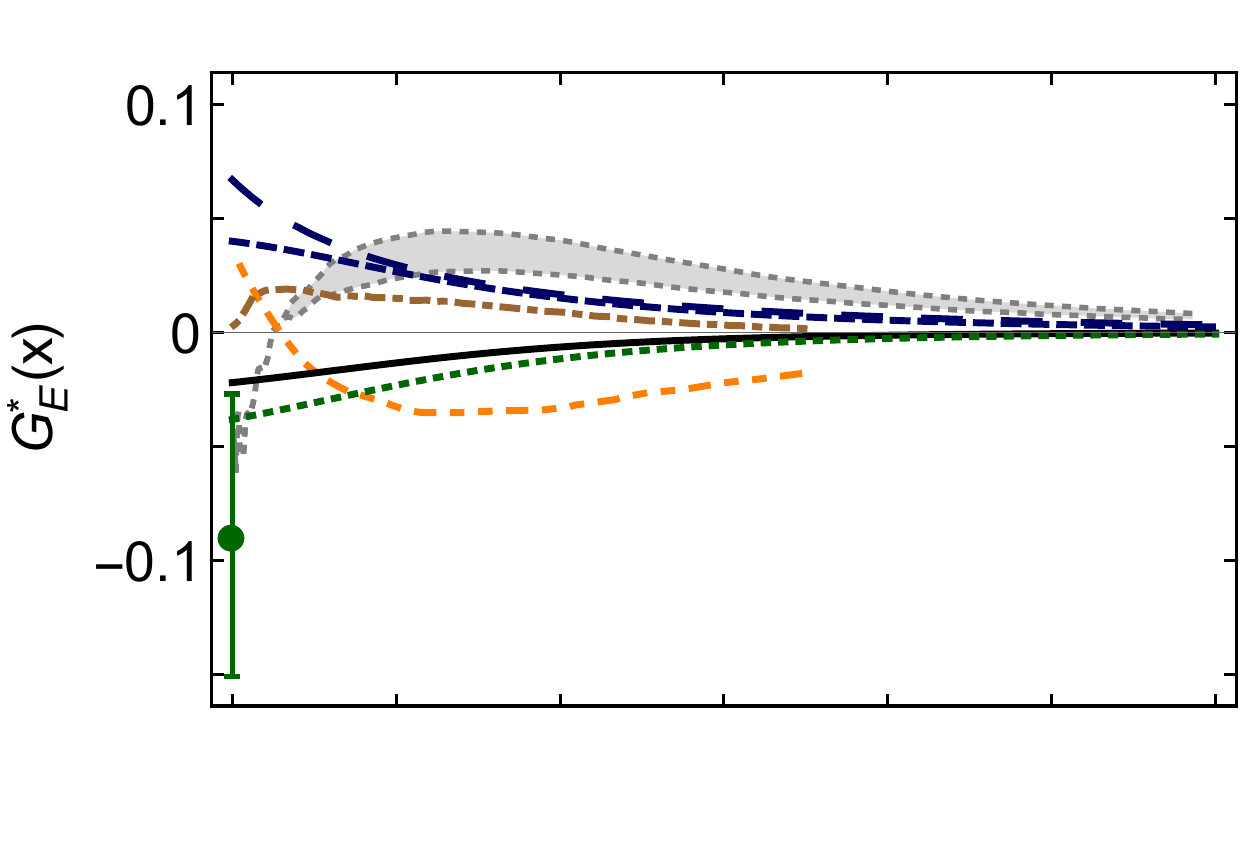}}
\vspace*{-8ex}

\centerline{%
\includegraphics[clip, width=0.42\textwidth]{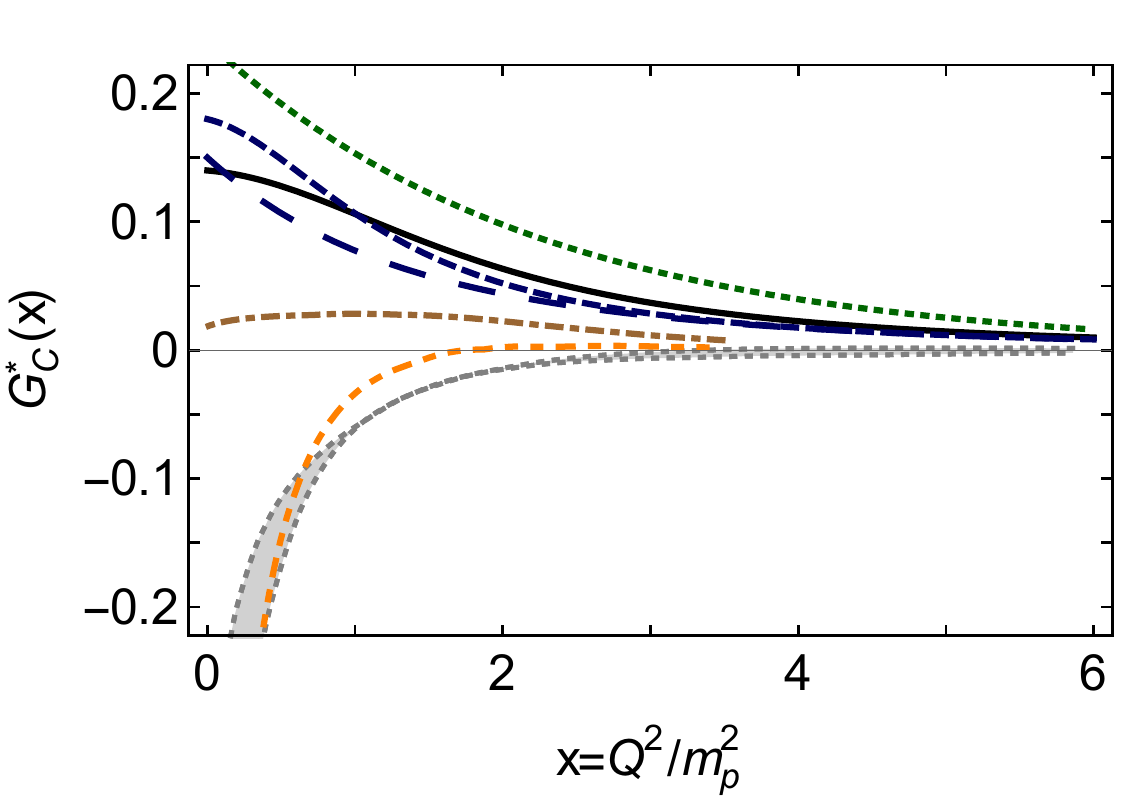}}
\caption{\label{D1600TFFs}
\emph{Top panel} -- Magnetic dipole $\gamma^\ast p\to \Delta^+(1600)$ transition form factor;
\emph{middle} -- electric quadrupole; and
\emph{bottom}:  Coulomb quadrupole.
Data from Ref.\,\cite{Tanabashi:2018oca}; and the conventions of Ref.\,\cite{Jones:1972ky} are employed.
All panels:
solid (black) curve, complete result;
long-dashed (blue) curve, result obtained when $\Delta(1600)$ is reduced to $S$-wave state;
dashed (blue) curve, both the proton and $\Delta(1600)$ are reduced to $S$-wave states;
dotted (green) curve, obtained by enhancing proton's axial-vector diquark content;
shaded (grey) band, light-front relativistic Hamiltonian dynamics (LFRHD) \cite{Capstick:1994ne};
dot-dashed (brown) curve, light-front relativistic quark model (LFRQM) with unmixed wave functions \cite{Aznauryan:2015zta};
and dot-dot-dashed (orange) curve, LFRQM with configuration mixing \cite{Aznauryan:2016wwm}.
}
\end{figure}

\section{Calculated Form Factors: $\mathbf {\Delta(1600)}$}
\label{SecTFFs1600}
Predictions for the $\gamma^\ast p\to \Delta^+(1600)$ transition form factors are displayed in Fig.\,\ref{D1600TFFs}.  Interpolations are provided by the simple functional form in Eq.\,\eqref{interpolationTFFs}, with the coefficients given in Table~\ref{TFFinterpolation}.  (Again, these forms should not be used for large-$x$ extrapolation.)  Empirical results are here only available at the real-photon point: $G_M^\ast(Q^2=0)$, $G_E^\ast(Q^2=0)$.  Evidently, the quark model results -- (shaded grey band) \cite{Capstick:1994ne}, dot-dashed (brown) curve \cite{Aznauryan:2015zta} and dot-dot-dashed (orange) curve \cite{Aznauryan:2016wwm}) -- are very sensitive to the wave functions employed for the initial and final states.  Furthermore, inclusion of relativistic effects has a sizeable impact on transitions to positive-parity excited states \cite{Capstick:1994ne}.

Our prediction is the solid (black) curve in each panel of Fig.\,\ref{D1600TFFs}.  In this instance, every transition form factor is of unique sign on the domain displayed.  Notably, the mismatches with the empirical results for $G_M^\ast(Q^2=0)$, $G_E^\ast(Q^2=0)$ are commensurate in relative sizes with those in the $\Delta(1232)$ case, suggesting that MB\,FSIs are of similar importance in both channels.

As remarked above, axial-vector diquark contributions interfere constructively with MB\,FSIs; hence, regarding form factors, one can mimic some effects of a meson cloud by modifying the axial-vector diquark content of the participating hadrons.  Accordingly, to illustrate the potential impact of MB\,FSIs, we computed the transition form factors using an enhanced axial-vector diquark content in the proton.  This was achieved by setting $m_{1^+} = m_{0^+} = 0.85\,$GeV, values with which the proton's mass is practically unchanged.  The procedure produced the dotted (green) curves in Fig.\,\ref{D1600TFFs}; better aligning the $x\simeq 0$ results with experiment and suggesting thereby that MB\,FSIs will improve our predictions.

The short-dashed (blue) curve in Fig.\,\ref{D1600TFFs} is the result obtained when only rest-frame $S$-wave components are retained in the wave functions of the proton and $\Delta(1600)$-baryon; and the long-dashed (blue) curve is that computed with a complete proton wave function and a $S$-wave-projected $\Delta(1600)$.  Once again, the higher partial-waves have a visible impact on all form factors, with $G_E^\ast$ being most affected: the higher waves produce a change in sign.
%% Need P+D in Delta to obtain change in sign.  Adding P to proton doesn't help much.
This reemphasises one of the conclusions from the quark model studies, \emph{viz}.\ data on the $\gamma^\ast p\to \Delta^+(1600)$ transition form factors will be sensitive to the structure of the $\Delta^+(1600)$.

\begin{figure}[t]
\centerline{%
\includegraphics[clip, width=0.42\textwidth]{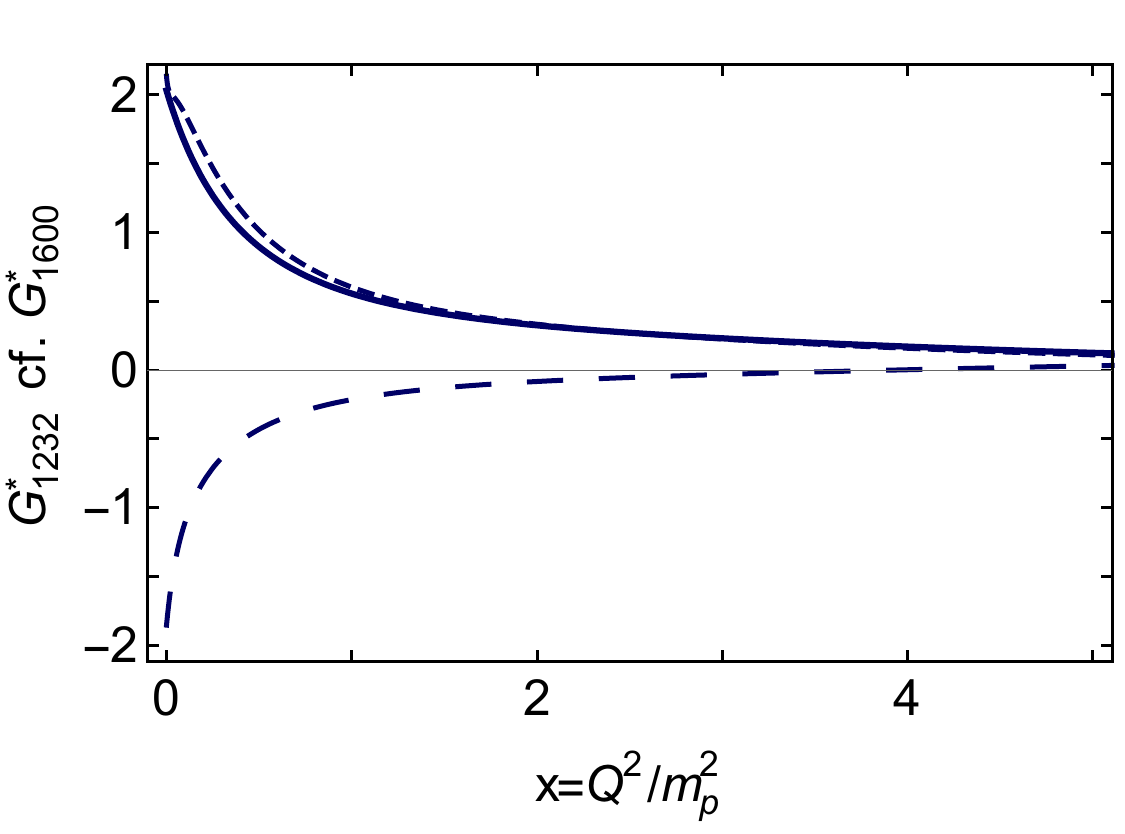}}
\caption{\label{D1232cfr1600}
Comparison between transition form factors via the following ratios:
solid curve, $\tfrac{1}{3} G_{M\,1232}^\ast/G_{M\,1600}^\ast$;
long-dashed curve, $G_{E\,1232}^\ast/G_{E\,1600}^\ast$;
short-dashed curve, $G_{C\,1232}^\ast/G_{C\,1600}^\ast$.
}
\end{figure}

A direct comparison between the $\gamma^\ast p\to \Delta^+(1232)$ and $\gamma^\ast p\to \Delta^+(1600)$ transition form factors is presented in Fig.\,\ref{D1232cfr1600}.  In all cases, the $\Delta(1232)$ form factors are larger in magnitude at small $x$.  However, with increasing $x$, there is always a point at which the ordering is reversed: $x\approx 2$ for $G_M^\ast$; $x\approx 0.5$ for $G_E^\ast$; and $x\approx 1$ for $G_C^\ast$.  These observations indicate that the dressed-quark-core component of the $\gamma^\ast p\to \Delta^+(1600)$ transition is more localised in configuration space, \emph{i.e}. more pointlike, than that of the $\gamma^\ast p\to \Delta^+(1232)$ transition.  In fact, using the dominant transition form factor, $G_M^\ast$, as a guide, the $\Delta^+(1600)$ transition radius is $\approx 1/3$ that of the $\Delta^+(1232)$.
%% Did Tiator and/or Vanderhaeghen study transverse charge densities for the Delta?

\begin{figure}[t]
\centerline{%
\includegraphics[clip, width=0.42\textwidth]{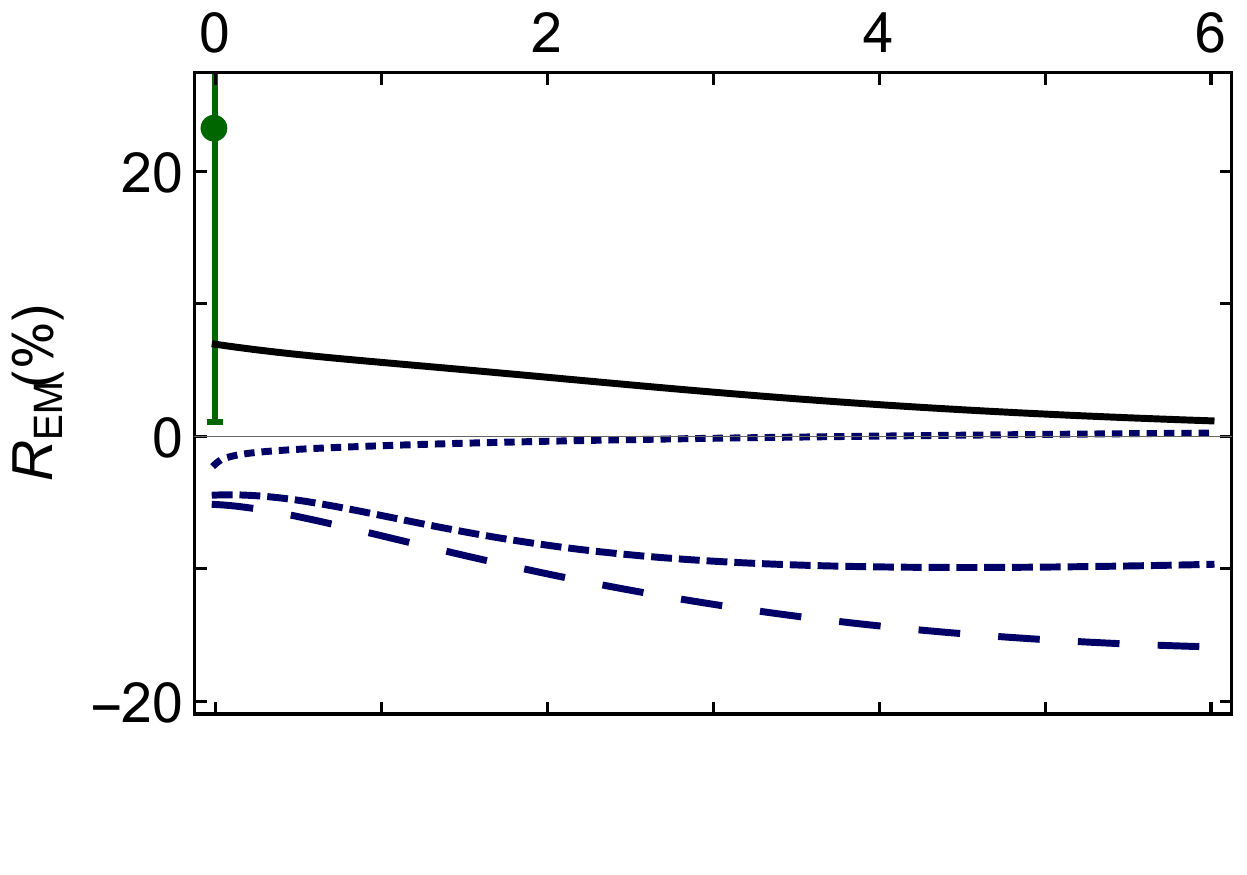}}
\vspace*{-9ex}

\centerline{%
\includegraphics[clip, width=0.42\textwidth]{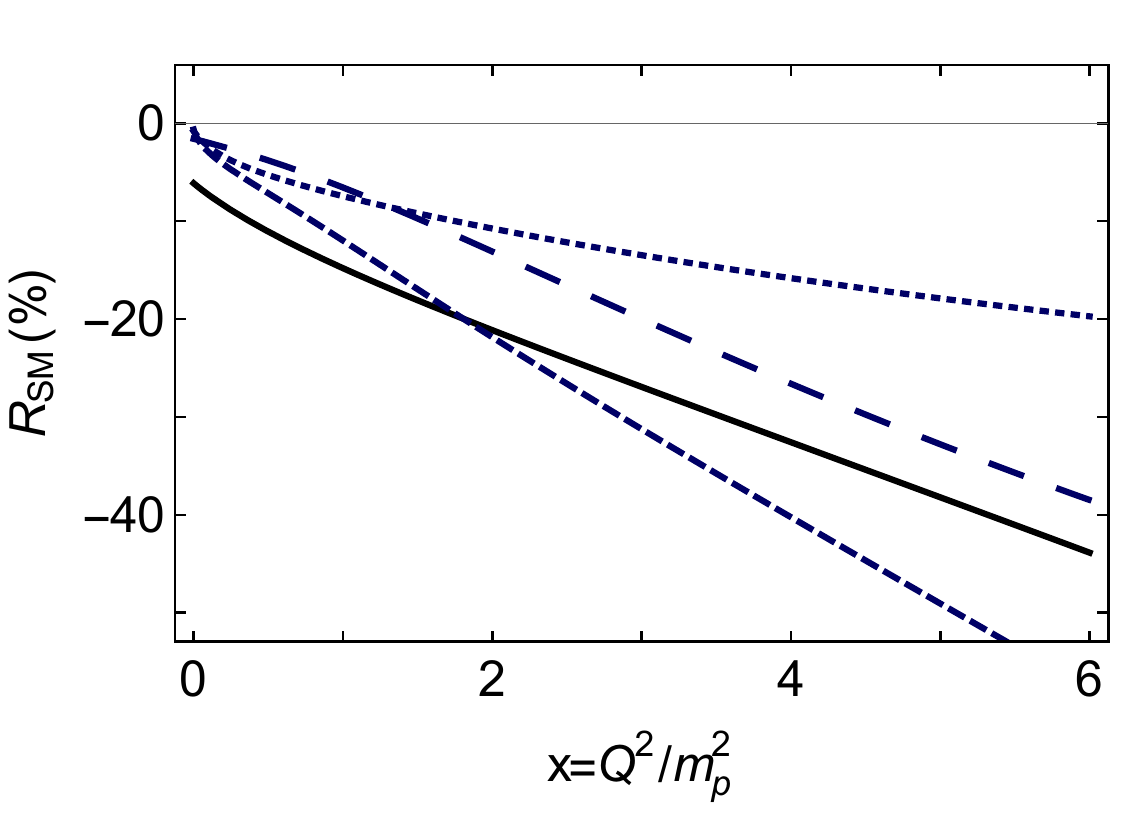}}

\caption{\label{figREMSM}
\emph{Top panel} -- $R_{EM}$.
Solid (black) curve -- our prediction for the $\gamma^\ast p \to \Delta^+(1600)$ transition;
long-dashed (blue) curve, result obtained when $\Delta(1600)$ is reduced to $S$-wave state;
dashed (blue) curve, obtained when both the proton and $\Delta(1600)$ are reduced to $S$-wave states;
dotted (blue) curve -- this ratio for $\gamma^\ast p \to \Delta^+(1232)$ transition.
\emph{Bottom panel} -- $R_{SM}$.  Legend as in the upper panel.
}
\end{figure}

Considering $\gamma^\ast p \to \Delta^+(1232)$, helicity conservation arguments within pQCD have been used to make the follow predictions for the ratios in Eq.\,\eqref{eqREMSM} \cite{Carlson:1985mm}:
\begin{equation}
\label{eqUVREMSM}
R_{EM} \stackrel{Q^2\to\infty}{=} 1 \,,\quad
R_{SM} \stackrel{Q^2\to\infty}{=} \,\mbox{\rm constant}\,,
\end{equation}
up to $\ln^2 Q^2$ corrections \cite{Idilbi:2003wj}.  These predictions disagree markedly with the outcomes produced by $SU(6)$-based quark models: $R_{EM} \equiv 0 \equiv R_{SM}$; and they are inconsistent with available data \cite{Aznauryan:2011ub, Aznauryan:2011qj}.  Notwithstanding such contradictions, Eqs.\,\eqref{eqUVREMSM} are indubitably correct, but evidence for approach to these limits will probably not become apparent until $x\gtrsim 20$ \cite{Segovia:2013uga}.

Our predictions for the ratios in Eqs.\,\eqref{eqREMSM} associated with the $\gamma^\ast p \to \Delta^+(1600)$ transition are depicted in Fig.\,\ref{figREMSM}.   The reasoning in Ref.\,\cite{Carlson:1985mm} should equally apply to this case; hence, Eqs.\,\eqref{eqUVREMSM} will become evident at some (very) large value of $x$.  At accessible scales, however, as we have repeatedly highlighted, dynamical features of the bound-state wave functions control the $x$-dependence of these ratios.  Examining Fig.\,\ref{figREMSM}, one sees that $R_{EM}$ for the $\Delta(1600)$ transition is far larger in magnitude than the analogous result for the $\Delta(1232)$ final state (and opposite in sign).  This is an observable manifestation of the enhanced $D$-wave strength in the $\Delta(1600)$ relative to that in the $\Delta(1232)$, which is apparent in Fig.\,\ref{10xistarS}.
%%
%% ground state ... 0+ -> 1+ dominant & S->P overlap
%% D1600 ... 1^+ -> 1+ equally important, which focuses on S->D ... 0+ -> 1+ & S->P has cancellation because of zeros in P-wave of D1600

\begin{figure*}[!t]
\begin{center}
\begin{tabular}{lr}
\includegraphics[clip,width=0.42\linewidth]{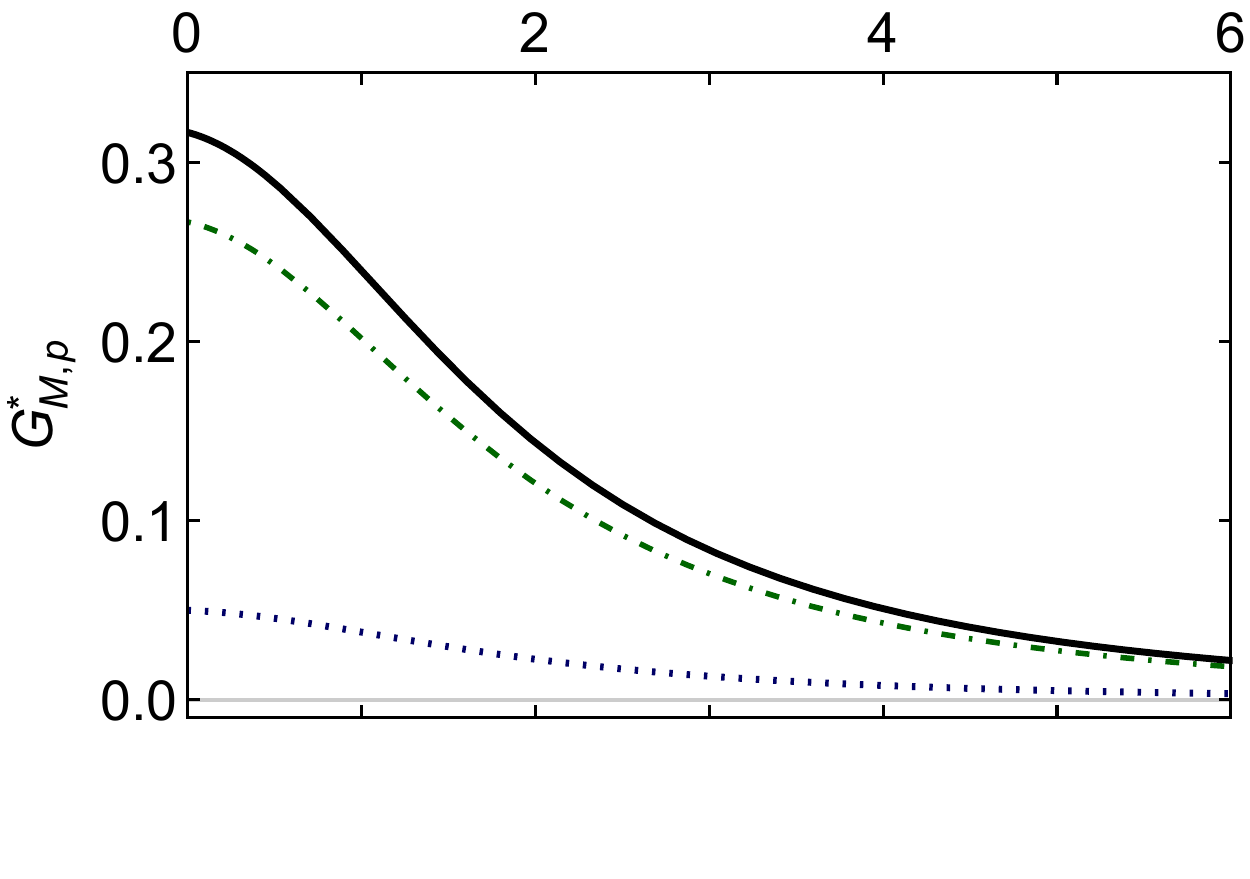}\hspace*{2ex } &
\includegraphics[clip,width=0.42\linewidth]{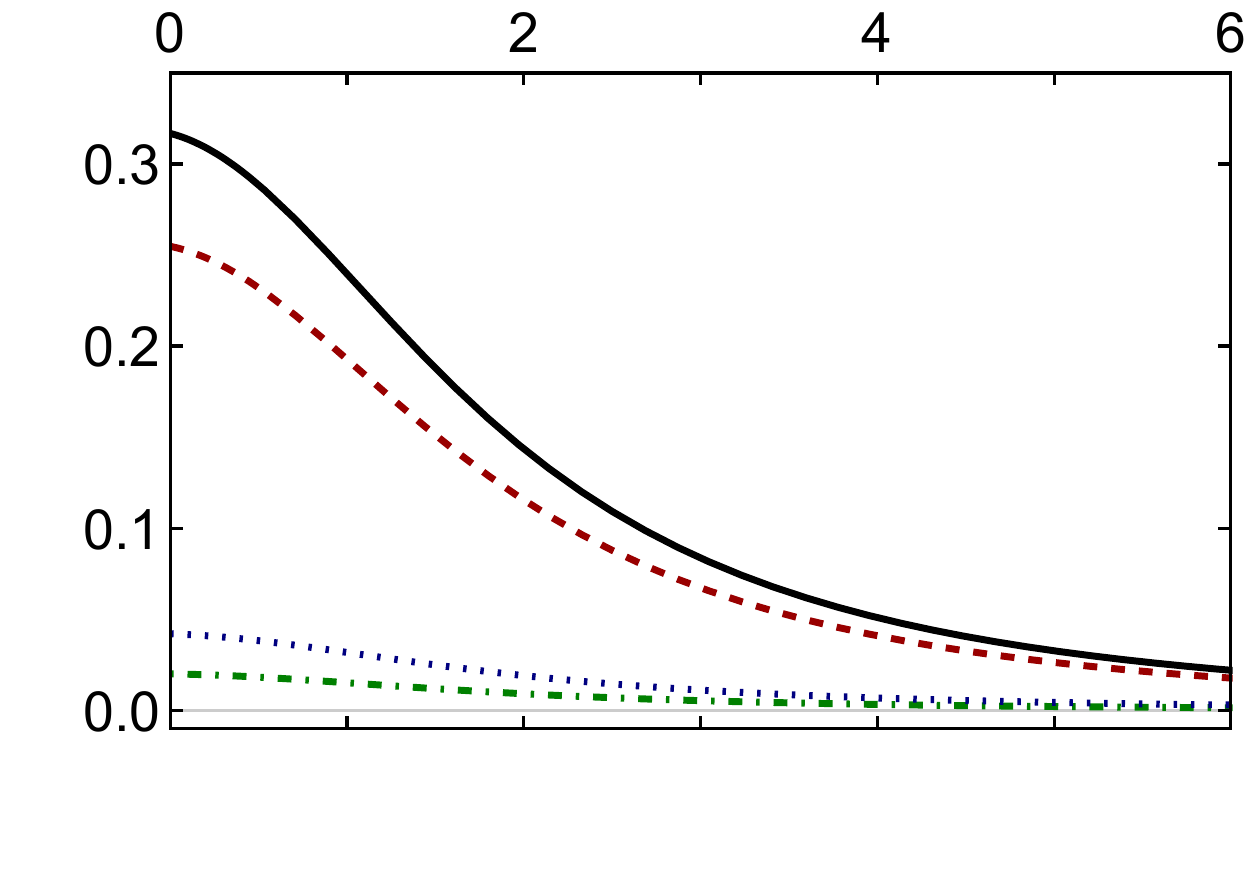}\vspace*{-8ex}
\end{tabular}
\begin{tabular}{lr}
\hspace*{-4ex}\includegraphics[clip,width=0.445\linewidth]{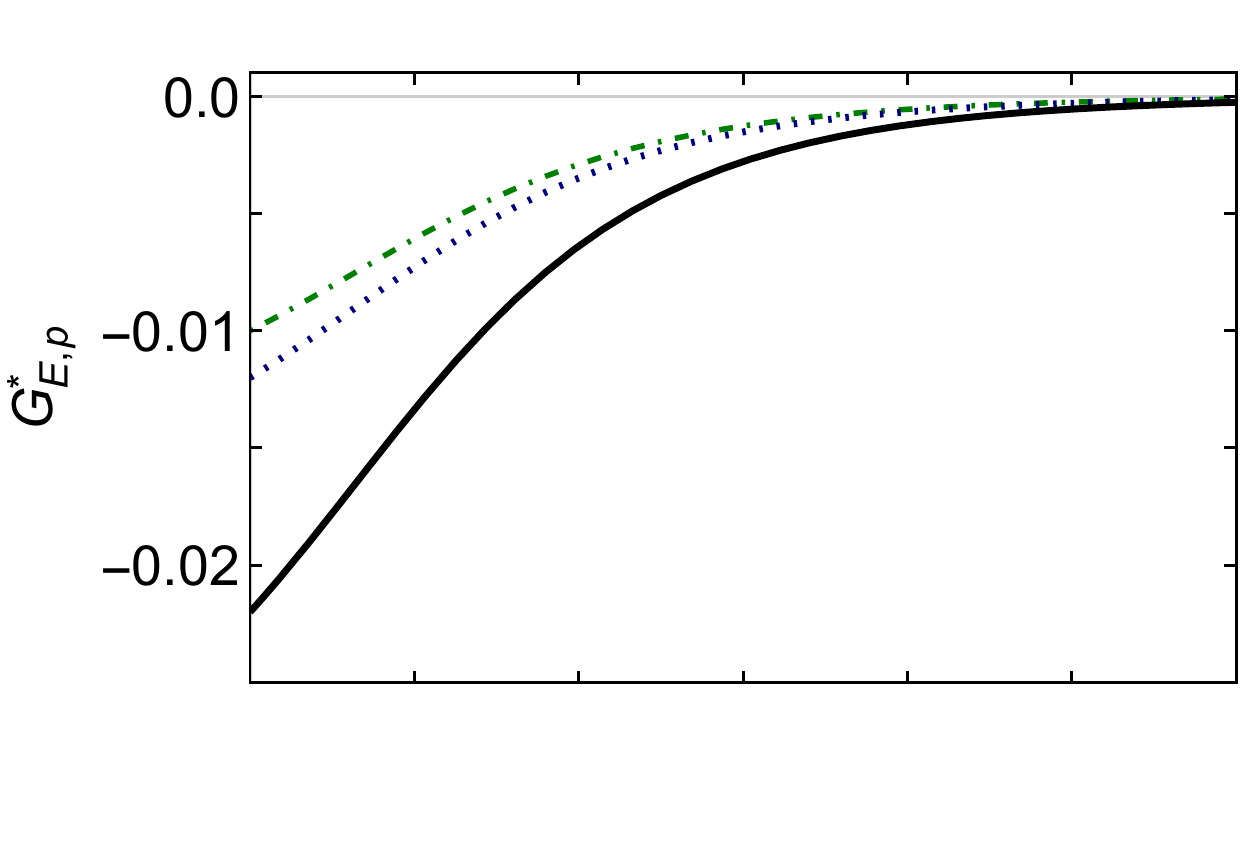}\hspace*{-1ex} &
\includegraphics[clip,width=0.44\linewidth]{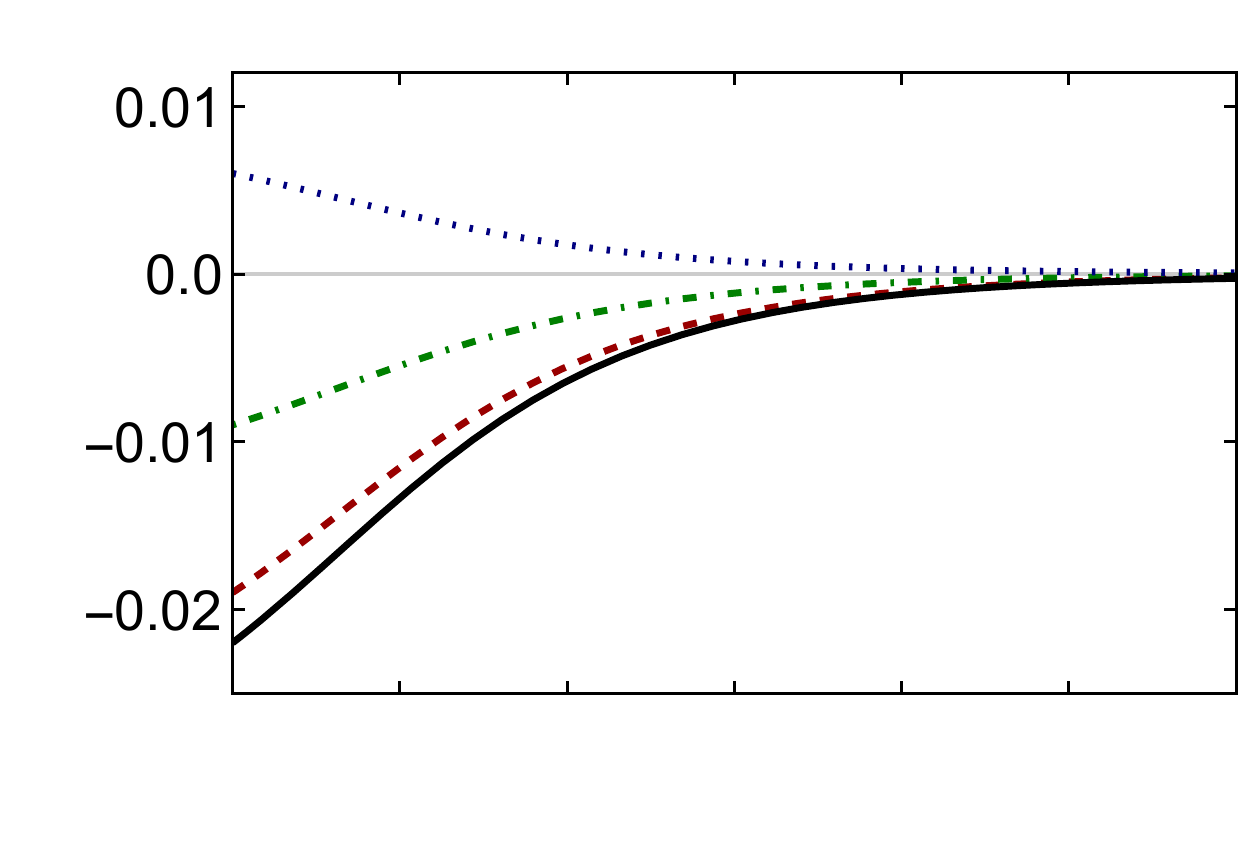}\vspace*{-8ex}
\end{tabular}
\begin{tabular}{lr}
\hspace*{-2ex}\includegraphics[clip,width=0.432\linewidth]{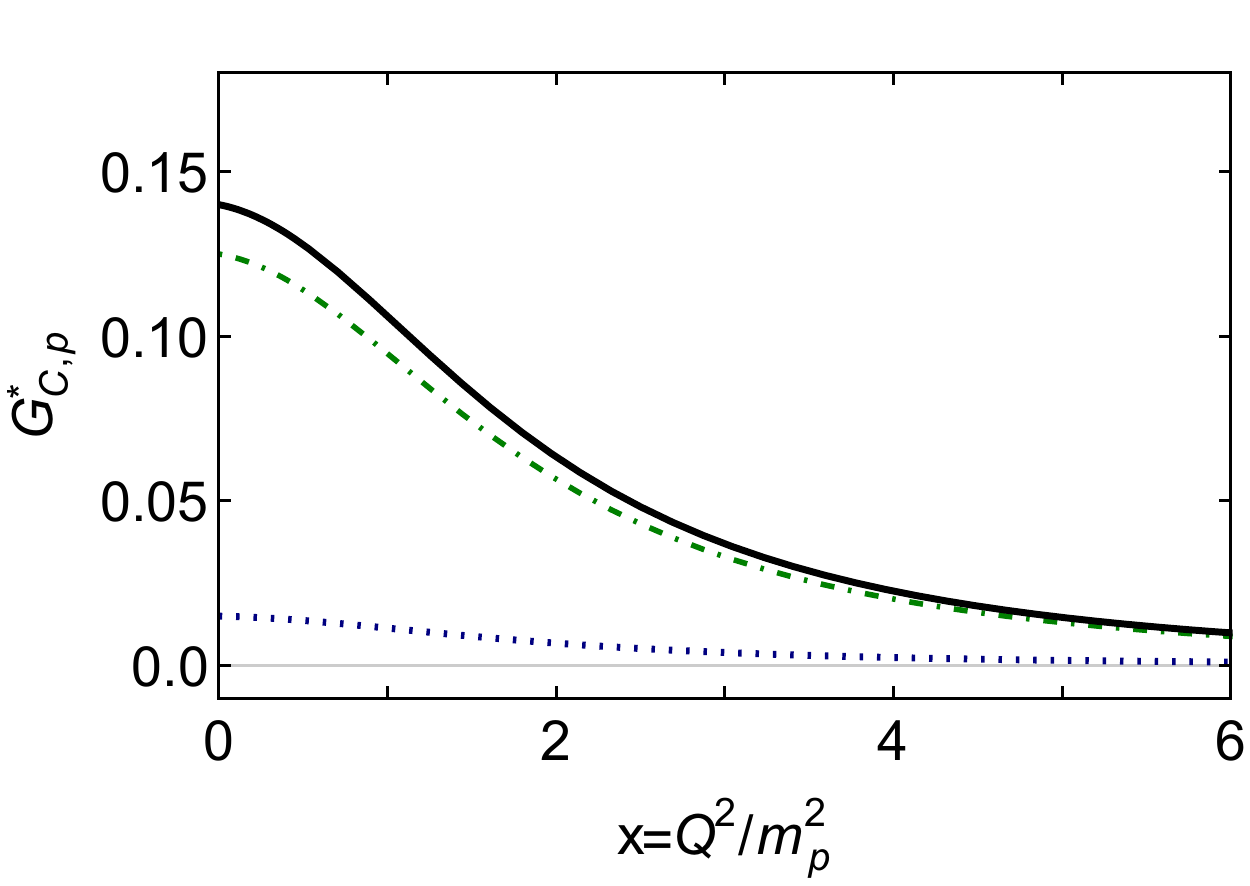}\hspace*{0.5ex } &
\includegraphics[clip,width=0.43\linewidth]{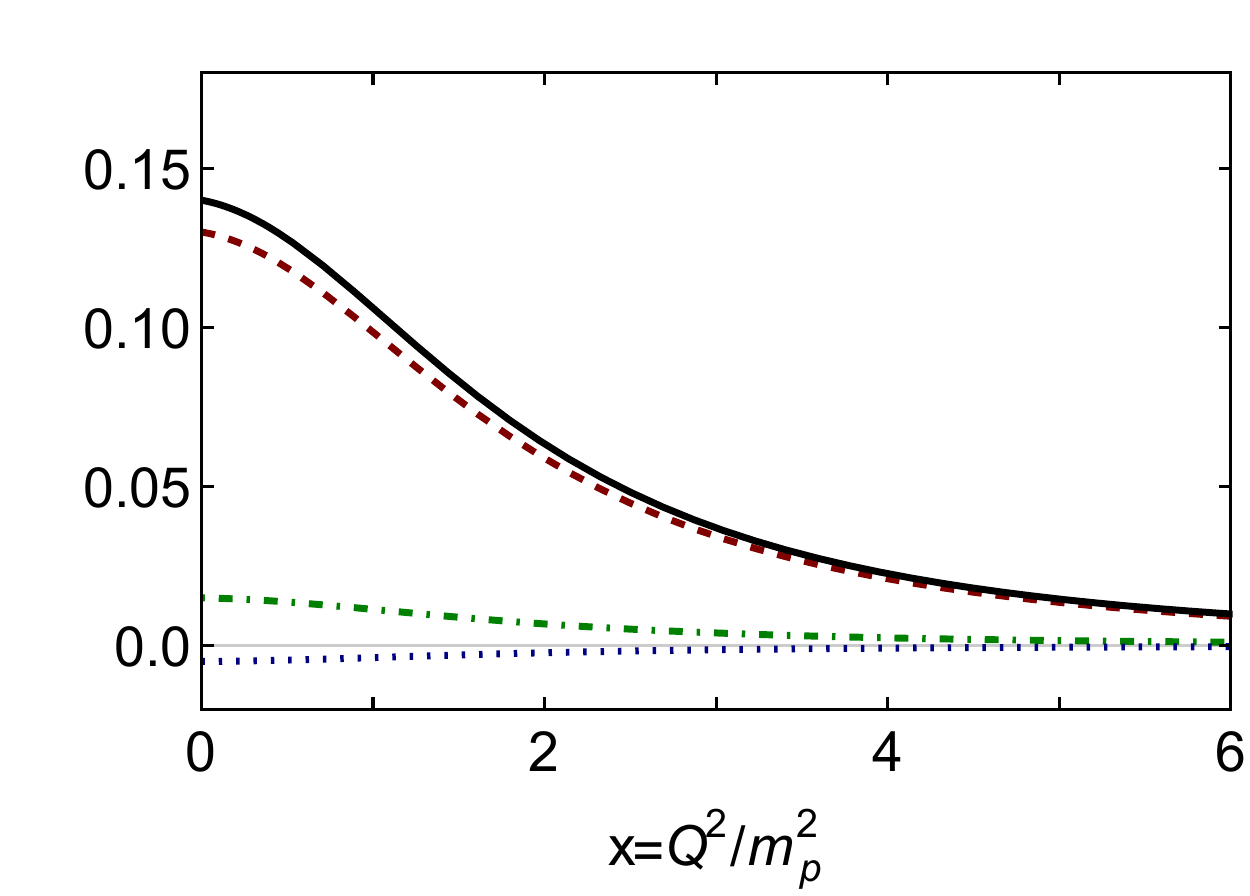}\vspace*{-0ex}
\end{tabular}
\end{center}
\caption{\label{Dissection}
$\gamma^\ast p \to \Delta^+(1600)$ transition form factors.
\emph{Left panels} -- diquark breakdown: \emph{DD2} (dot-dashed green), pseudovector diquark in both initial and final states; \emph{DD3} (dotted blue), scalar diquark in incoming baryon, pseudovector diquark in outgoing baryon.
\emph{Right panels} -- scatterer breakdown: \emph{DS1} (red dashed), photon strikes an uncorrelated dressed quark; \emph{DS2} (dot-dashed green), photon strikes a diquark; and \emph{DS3} (dotted blue), diquark breakup contributions, including photon striking exchanged dressed-quark.
}
\end{figure*}

\section{Form Factor Dissections: $\mathbf {\Delta(1600)}$}
\label{1600Dissections}
In connection with Eq.\,\eqref{JDeltaNExplicit}, we noted that the vertex sufficient to express the interaction of a photon with a baryon generated by the Faddeev equation in Fig.\,\ref{figFaddeev} is a sum of six terms, with the photon separately probing the quarks and diquarks in various ways.  Hence, diverse features of quark dressing and the quark-quark correlations all play a role in determining the form factors.  To elaborate, electroproduction form factors involving the nucleon and its excitations may be dissected in two separate ways, each of which can be considered as a sum of three distinct terms \cite{Segovia:2016zyc}.
\begin{description}
\item[DD = diquark dissection]$\,$\\[-3.5ex]
\begin{description}
\setlength\itemsep{0em}
\item[\mbox{\emph{DD1}}] sca\-lar diquark, $[ud]$, in both the initial- and final-state baryon,
%%$[ud]$, is a larger part of the nucleon's Faddeev amplitude, contributions with a pseudovector diquark, $\{qq\}$, in both $p$ and $\Delta^+$ contribute more strongly to the transition.
%
\item[\emph{DD2}] pseudovector diquark in both the initial- and final-state ($\Delta^+$: $\{uu\}$ or $\{ud\}$), and
\item[\emph{DD3}] a different diquark in the initial- and final-state.
\end{description}
\item[DS = scatterer dissection]$\,$\\[-3.5ex]
\begin{description}
\setlength\itemsep{0em}
\item[\emph{DS1}] photon strikes a bystander dressed-quark, with the accompanying diquark untouched (Diagram~1 in Ref.\,\cite{Segovia:2014aza}, Fig.C.1);
\item[\emph{DS2}] photon interacts with a diquark, elastically or causing a transition scalar\,$\leftrightarrow$\,pseudovector whilst the accompanying bystander quark is unaffected (Diagrams~2 and 4 in Ref.\,\cite{Segovia:2014aza}, Fig.C.1); and
\item[\emph{DS3}] photon strikes a dressed-quark in-flight, as one diquark breaks up and another is formed (Diagram ~3 in Ref.\,\cite{Segovia:2014aza}, Fig.C.1), or appears in one of the two associated ``seagull'' terms (Diagrams~5 and 6).
\end{description}
\end{description}
%Here, the ``bystander'' quark is that which is not within the quark-quark correlation.
The anatomy of a given transition is revealed by merging the information provided by DD and DS.  With a $\Delta$-baryon in the final state, \emph{DD1} does not contribute because the $I=0$ diquark plays no role in an $I=3/2$ baryon.

The structure of $G_{M}^{\ast}$ in the $\gamma^\ast p \to \Delta(1600)$ transition is revealed in the upper row of Fig.\,\ref{Dissection}.  The left panel shows that \emph{DD2} is far stronger than \emph{DD3} and the right panel reveals that \emph{DS1} is overwhelmingly dominant, \emph{viz}.\ the largest contribution to $G_{M}^{\ast}$ is provided by diagrams in which a photon scatters from the bystander quark, flipping its spin, in the presence of an idle pseudovector diquark.
This is similar to the nature of $G_{M}^{\ast}$ in the $\gamma^\ast p \to \Delta(1232)$ transition \cite{Segovia:2016zyc}, although the $0^+$-to-$1^+$ diquark transition component is a much smaller fraction for the $\Delta(1600)$ final state. One should also recall Fig.\,\ref{D1232cfr1600}, which depicts the $x$-dependence of the relative magnitudes of $G_{M}^{\ast}$ for the two final states.
%% GM D1600 much smaller at small x ... but becomes larger on x>2

The electric quadrupole transition form factor, $G_{E}^{\ast}$, for the $\Delta(1600)$ final state is dissected in the middle row of Fig.\,\ref{Dissection}.
The left panel shows that \emph{DD2} and \emph{DD3} are of comparable size; and the right panel, that \emph{DS1} is dominant, with \emph{DS2} and \emph{DS3} approximately cancelling.  Hence, the transition is dominated by diagrams in which the photon scatters from the bystander quark, leaving its spin unchanged, with the strength of the transition resulting from the overlap between what may be said to be quark-diquark components in the rest-frame Faddeev wave functions of the proton and $\Delta^+(1600)$ that differ by two units of angular momentum.
This is markedly different from $G_{E}^{\ast}$ in the $\Delta(1232)$ transition, described in connection with Fig.\,\ref{D1600TFFs}.

The anatomy of $G_{C}^{\ast}$ is revealed in the bottom row of Fig.\,\ref{Dissection}.  Evidently, the behaviour is largely determined by \emph{DD2} and \emph{DS1} processes, \emph{i.e}.\ the transition strength and $x$-dependence measure the overlap between $S$- and $D$-wave quark-diquark angular momentum components in the rest-frame proton and $\Delta^+(1600)$ Faddeev wave functions.
%\textcolor[rgb]{1.00,0.00,0.00}{$\chi$-dependence of $G_C^\ast$}
% Small ... -15% if one turns it off
Much the same is true in the $\gamma^\ast p \to \Delta(1232)$ transition \cite{Segovia:2016zyc}.

\section{Summary and Perspectives}
\label{epilogue}
We computed $\gamma^\ast p \to \Delta^+(1232), \Delta^+(1600)$ transition form factors using a quark-diquark approximation to the Poincar\'e-covariant three-body bound-state problem in relativistic quantum field theory, unifying their treatment with that of nucleon elastic form factors \cite{Segovia:2014aza} and $\gamma^\ast N \to R$ transitions \cite{Segovia:2015hra, Chen:2018nsg}.  Crucially, the diquark correlations are nonpointlike and fully-dynamical, and the Faddeev kernel ensures that every valence-quark participates actively in all diquark correlations to the fullest extent allowed by kinematics and symmetries.  Moreover, each dressed-quark is characterised by a nonperturbatively generated running mass function, expressing a signature consequence of dynamical chiral symmetry breaking in the Standard Model \cite{NAP13438, Brodsky:2015aia}.  The $\Delta(1600)$-baryon generated by this approach is the simplest radial excitation of the $\Delta(1232)$ (Sec.\,\ref{SecFaddeev}, Fig.\,\ref{10xistarS}), \emph{viz}.\ it is analogous to the Roper resonance in the nucleon sector \cite{Burkert:2017djo}.

Regarding the $\gamma^\ast p \to \Delta^+(1232)$ transition, precise measurements already exist on $0 \leq Q^2 \lesssim 8\,$GeV$^2$ \cite{Aznauryan:2011ub, Aznauryan:2011qj}; and on $Q^2 \gtrsim 0.5 m_p^2$, \emph{i.e}.\ outside the meson cloud domain for this process, our calculated magnetic dipole and Coulomb quadrupole form factors agree well with this data (Sec.\,\ref{SecTFFs1232}, Figs.\,\ref{figD1232}, \ref{figD1232ratios}).  Consistent with the data, too, we find that the electric quadrupole form factor is very small in magnitude; hence, it is particularly sensitive to the diquark content and quark-diquark angular-momentum structure of the baryons involved, and also to meson-baryon final-state-interactions (MB\,FSIs) on a larger domain than the other form factors.

Our predictions for the $\gamma^\ast p \to \Delta^+(1600)$ magnetic dipole and electric quadrupole transition form factors are consistent with the empirical values at the real photon point, but we expect inclusion of meson-baryon final-state-interactions to improve the agreement on $Q^2\simeq 0$ (Sec.\,\ref{SecTFFs1600}, Fig.\,\ref{D1600TFFs}).
On the other hand, the predictions extend to $Q^2 \approx 6 m_p^2$, \emph{i.e}.\ beyond the meson-cloud domain; hence, a meaningful direct comparison with existing data \cite{Trivedi:2018rgo, Burkert:2019opk} will be possible once the analysis is completed.

It is interesting to observe that whilst all $\gamma^\ast p \to \Delta^+(1232)$ transition form factors are larger in magnitude than those for $\gamma^\ast p \to \Delta^+(1600)$ in some neighbourhood of $Q^2=0$, this ordering is reversed on $Q^2 \gtrsim 2 m_p^2$ (Sec.\,\ref{SecTFFs1600}, Fig.\,\ref{D1232cfr1600}).  One can thus argue that the $\gamma^\ast p \to \Delta^+(1600)$ transition is more localised in configuration space.

It is also notable that $R_{SM}$ is qualitatively similar for both transitions considered herein; but $R_{EM}$ is markedly different, being of opposite sign on $Q^2 \lesssim 4 m_p^2$ and uniformly larger in magnitude for the $\Delta(1600)$ (Sec.\,\ref{SecTFFs1600}, Fig.\,\ref{figREMSM}).  These observations again highlight the sensitivity of the electric quadrupole form factor to the degree of deformation of the $\Delta$-baryons.  Diquark and scatterer dissections of the transition form factors were useful in developing an understanding of the key reaction mechanisms for each electroproduction form factor  (Sec.\,\ref{1600Dissections}).

There are numerous worthwhile extensions of our analysis, \emph{e.g}.\ calculation of the $\gamma^\ast p \to N(1535)\,1/2^-$ transition form factors is already underway.  Here the final-state is the nucleon's parity-partner, which holds a special place in QCD owing to the manifest role of dynamical chiral symmetry breaking (DCSB) in generating the mass-splitting between this state and the nucleon.  The $N(1535)\,1/2^-$ wave function is qualitatively different to that of the near-lying Roper resonance \cite{Chen:2017pse}: the pointwise behaviour of each component is simpler, but there are more components because pseudoscalar and vector diquark correlations are also present in this negative-parity bound-state.  Consequently, analyses of $\gamma^\ast p \to N(1535)\,1/2^-$ explore novel aspects of baryon structure.  For instance, as $\gamma^\ast p \to\Delta$ transitions are sensitive to the relative strength of scalar and axial-vector diquarks within the proton  (Sec.\,\ref{1600Dissections}), then one should expect $\gamma^\ast p \to N(1535)\,1/2^-$ to reveal the relative strength of positive and negative parity diquarks in the $N(1535)\,1/2^-$ because negative-parity diquarks are negligible in the proton.  Any predictions one makes can immediately be tested because data exists on $0\leq Q^2 \lesssim 6 m_p^2$ \cite{Aznauryan:2011ub, Aznauryan:2011qj}.  Moreover, given the relative ease of separating low-lying states of opposite parity, lattice-regularised QCD may also be able to contribute \cite{Braun:2009jy}.

An analogue of $\gamma^\ast p \to N(1535)\,1/2^-$ is $\gamma^\ast p \to \Delta(1700) 3/2^-$, in which the final state is the $\Delta$-baryon's parity partner.  Comparison between the electroproduction form factors for this process and those calculated herein would provide additional insights into the role played by DCSB in hadron structure.

Computation of $\gamma^\ast p \to N(1710)\,1/2^+$ electroproduction form factors is also valuable because the structure of the $N(1710)\,1/2^+$ is unclear.  In quark models, the profile of its wave function is sensitive to the formulation.  For instance, it can be Roper-like, with two peaks skewed relative to those in the kindred Roper wave function \cite{Capstick:1992xn, Melde:2008yr}, in which case it may be a candidate for the system which is predominantly quark-plus-radially-excited-diquark; or it can have three peaks, located on the same trajectory as the two in the related Roper wave function \cite{Santopinto:2012nq, deTeramond:2014asa}, \emph{viz}.\ the second radial excitation of the quark-plus-diquark system.  A third possibility, realised in some dynamical coupled channels (DCC) calculations \cite{Suzuki:2009nj}, sees the Roper and $N(1710)\,1/2^+$ as both derived from the same quark core state.  Given that $N(1710)\,1/2^+$ electroproduction data exist on $Q^2\lesssim 4\,m_p^2$ \cite{Park:2014yea} and that each helicity amplitude appears to be of unique sign, unlike those for the Roper \cite{Aznauryan:2009mx, Mokeev:2012vsa, Mokeev:2015lda}, it is worth testing these possibilities by exploring the solution space of the Poincar\'e-covariant Faddeev equation and using the results to compute the transition form factors.

As a final class of examples, we note that a complement to the analyses highlighted above is offered by studies of electroproduction form factors for low-lying baryons with ``mixed'' spin-isospin structure, \emph{viz}.\ $(I,J)=(1/2,3/2^\pm), (3/2,1/2^\pm)$.  For such systems, the normal level-ordering has negative-parity states lighter than positive-parity states: DCSB must still generate the (large) splitting from the ground state baryon, but the connection with parity is reversed.
%% gamma_5 goes with the positive parity systems because QMs give L=1 for negative-parity and need gamma_5 to flip it.
%% D13 +/- parity N1720 3/2+ and N1520 3/2- ... D13(1520)
%% I=3/2, J=1/2 +/- Delta(1910) 1/2+ Delta(1620) 1/2 -
%%
Data on the $N(1520) 3/2^-$ electrocouplings are available to $Q^2\lesssim 4 m_p^2$ \cite{Dugger:2009pn, Aznauryan:2009mx, Mokeev:2012vsa, Mokeev:2015lda}.  Based upon this, some coupled-channels studies indicate that MB\,FSIs are (almost) negligible for the $A_{1/2}$ helicity amplitude \cite{JuliaDiaz:2007fa, Aznauryan:2012ba}, the calculation of which might therefore serve as a good test of the dressed-quark-core approach exploited herein.

\bigskip

\begin{acknowledgments}
%\centerline{\textbf{ACKNOWLEDGMENTS}}

%\smallskip

We are grateful for constructive comments from \mbox{L.~Chang}, R.~Gothe, V.~Mokeev, K.~Raya, F.~Wang and S.-S.~Xu;
for the hospitality and support of RWTH Aachen University, III.\,Physikalisches Institut B, Aachen - Germany;
and likewise for the hospitality and support of the University of Huelva, Huelva - Spain, and the University of Pablo de Olavide, Seville - Spain, during the ``4th Workshop on Nonperturbative QCD'' (University of Pablo de Olavide, 6-9 November 2018).
Work supported by:
National Natural Science Foundation of China, grant nos.\ 11535005, 11690030, 11805097;
Jiangsu Province Natural Science Foundation grant no.\ BK20180323;
Jiangsu Province \emph{Hundred Talents Plan for Professionals};
Funda\c{c}\~ao de Amparo \`a Pesquisa do Estado de S\~ao Paulo - FAPESP grant no.\,2015/21550-4;
U.S.\ Department of Energy, Office of Science, Office of Nuclear Physics, under contract no.\,DE-AC02-06CH11357;
Forschungszentrum J\"ulich GmbH;
and Ministerio de Economia Industria y Competitividad (MINECO), under grant no.\,FPA2017-86380-P.
\end{acknowledgments}

%%%%

%%\bibliographystyle{../../../zProc/z10/z10KITPC/h-physrev4}
%%\bibliography{../../../CollectedBiB}

\begin{thebibliography}{100}

\bibitem{Fermi:1952zz}
E.~Fermi, H.~Anderson, A.~Lundby, D.~Nagle and G.~Yodh,
\newblock Phys. Rev. {\bf 85}, 935 (1952).
%%CITATION = PHRVA,85,935;%%

\bibitem{Anderson:1952nw}
H.~Anderson, E.~Fermi, E.~Long and D.~Nagle,
\newblock Phys. Rev. {\bf 85}, 936 (1952).
%%CITATION = PHRVA,85,936;%%

\bibitem{Nagle:1984sg}
D.~E. Nagle,
\newblock (1984),
\newblock {\emph{The Delta: The First Pion Nucleon Resonance, Its Discovery and
  Applications}, Los Alamos National Laboratory Report no. LALP-84-27}.
%%CITATION = LALP-84-27 ETC.;%%

\bibitem{Tanabashi:2018oca}
M.~Tanabashi {\em et~al.},
\newblock Phys. Rev. D {\bf 98}, 030001 (2018),
\newblock {(\emph{Particle Data Group})}.
%%CITATION = PHRVA,D98,030001;%%

\bibitem{Aznauryan:2011ub}
I.~Aznauryan, V.~Burkert, T.-S. Lee and V.~Mokeev,
\newblock J. Phys. Conf. Ser. {\bf 299}, 012008 (2011).
%%CITATION = ARXIV:1102.0597;%%

\bibitem{Aznauryan:2011qj}
I.~Aznauryan and V.~Burkert,
\newblock Prog. Part. Nucl. Phys. {\bf 67}, 1 (2012).
%%CITATION = ARXIV:1109.1720;%%

\bibitem{Aznauryan:2012ba}
I.~Aznauryan {\em et~al.},
\newblock Int. J. Mod. Phys. E {\bf 22}, 1330015 (2013).
%%CITATION = ARXIV:1212.4891;%%

\bibitem{Carlson:1985mm}
C.~E. Carlson,
\newblock Phys. Rev. D {\bf 34}, 2704 (1986).
%%CITATION = PHRVA,D34,2704;%%

\bibitem{Pascalutsa:2006up}
V.~Pascalutsa, M.~Vanderhaeghen and S.~N. Yang,
\newblock Phys. Rept. {\bf 437}, 125 (2007).
%%CITATION = HEP-PH/0609004;%%

\bibitem{Eichmann:2011aa}
G.~Eichmann and D.~Nicmorus,
\newblock Phys. Rev. D {\bf 85}, 093004 (2012).
%%CITATION = ARXIV:1112.2232;%%

\bibitem{Segovia:2013rca}
J.~Segovia, C.~Chen, C.~D. Roberts and S.-L. Wan,
\newblock Phys. Rev. C {\bf 88}, 032201(R) (2013).
%%CITATION = ARXIV:1305.0292;%%

\bibitem{Segovia:2013uga}
J.~Segovia {\em et~al.},
\newblock Few Body Syst. {\bf 55}, 1 (2014).
%%CITATION = ARXIV:1308.5225;%%

\bibitem{Segovia:2014aza}
J.~Segovia, I.~C. Clo{\"e}t, C.~D. Roberts and S.~M. Schmidt,
\newblock Few Body Syst. {\bf 55}, 1185 (2014).
%%CITATION = ARXIV:1408.2919;%%

\bibitem{Segovia:2016zyc}
J.~Segovia and C.~D. Roberts,
\newblock Phys. Rev. C {\bf 94}, 042201(R) (2016).
%%CITATION = ARXIV:1607.04405;%%

\bibitem{Alexandrou:2012da}
C.~Alexandrou, C.~Papanicolas and M.~Vanderhaeghen,
\newblock Rev. Mod. Phys. {\bf 84}, 1231 (2012).
%%CITATION = ARXIV:1201.4511;%%

\bibitem{Santopinto:2012nq}
E.~Santopinto and M.~M. Giannini,
\newblock Phys. Rev. C {\bf 86}, 065202 (2012).
%%CITATION = ARXIV:1506.01207;%%

\bibitem{Sanchis-Alepuz:2017mir}
H.~Sanchis-Alepuz, R.~Alkofer and C.~S. Fischer,
\newblock Eur. Phys. J. A {\bf 54}, 41 (2018).
%%CITATION = ARXIV:1707.08463;%%

\bibitem{Buchmann:2018nmu}
A.~J. Buchmann,
\newblock Few Body Syst. {\bf 59}, 145 (2018).
%%CITATION = ARXIV:1902.10166;%%

\bibitem{Roberts:2015dea}
C.~D. Roberts,
\newblock J. Phys. Conf. Ser. {\bf 630}, 012051 (2015).
%%CITATION = ARXIV:1501.06581;%%

\bibitem{Burkert:2018oyl}
V.~D. Burkert,
\newblock Few Body Syst. {\bf 59}, 57 (2018).
%%CITATION = ARXIV:1801.10480;%%

\bibitem{Roberts:2018hpf}
C.~D. Roberts,
\newblock Few Body Syst. {\bf 59}, 72 (2018).
%%CITATION = ARXIV:1801.08562;%%

\bibitem{Roper:1964zza}
L.~D. Roper,
\newblock Phys. Rev. Lett. {\bf 12}, 340 (1964).
%%CITATION = PRLTA,12,340;%%

\bibitem{BAREYRE1964137}
P.~Bareyre {\em et~al.},
\newblock Physics Letters {\bf 8}, 137 (1964).

\bibitem{AUVIL196476}
P.~Auvil, C.~Lovelace, A.~Donnachie and A.~Lea,
\newblock Physics Letters {\bf 12}, 76 (1964).

\bibitem{PhysRevLett.13.555}
S.~L. Adelman,
\newblock Phys. Rev. Lett. {\bf 13}, 555 (1964).

\bibitem{PhysRev.138.B190}
L.~D. Roper, R.~M. Wright and B.~T. Feld,
\newblock Phys. Rev. {\bf 138}, B190 (1965).

\bibitem{Capstick:2000qj}
S.~Capstick and W.~Roberts,
\newblock Prog. Part. Nucl. Phys. {\bf 45}, S241 (2000).
%%CITATION = NUCL-TH/0008028;%%

\bibitem{Crede:2013sze}
V.~Crede and W.~Roberts,
\newblock Rept. Prog. Phys. {\bf 76}, 076301 (2013).
%%CITATION = ARXIV:1302.7299;%%

\bibitem{Giannini:2015zia}
M.~M. Giannini and E.~Santopinto,
\newblock Chin. J. Phys. {\bf 53}, 020301 (2015), [1501.03722].
%%CITATION = ARXIV:1501.03722;%%

\bibitem{Golli:2017nid}
B.~Golli, H.~Osmanovi{\'c}, S.~{\v{S}}irca and A.~{\v{S}}varc,
\newblock Phys. Rev. C {\bf 97}, 035204 (2018).
%%CITATION = ARXIV:1709.09025;%%

\bibitem{Burkert:2017djo}
V.~D. Burkert and C.~D. Roberts,
\newblock Rev. Mod. Phys. {\bf 91}, 011003 (2019).
%%CITATION = ARXIV:1710.02549;%%

\bibitem{Trivedi:2018rgo}
A.~Trivedi,
\newblock Few Body Syst. {\bf 60}, 5 (2019).
%%CITATION = FBSYE,60,;%%

\bibitem{Burkert:2019opk}
V.~D. Burkert, V.~I. Mokeev and B.~S. Ishkhanov,
\newblock (arXiv:1901.09709 [nucl-ex]),
\newblock {\emph{The nucleon resonance structure from exclusive $\pi^+\pi^-p$
  photo-/electroproduction off protons}}.
%%CITATION = ARXIV:1901.09709;%%

\bibitem{Chen:2019fzn}
C.~Chen, G.~Krein, C.~D. Roberts, S.~M. Schmidt and J.~Segovia,
\newblock (arXiv:1901.04305 [nucl-th]),
\newblock {\emph{Spectrum and structure of octet and decuplet baryons and their
  positive-parity excitations}}.
%%CITATION = ARXIV:1901.04305;%%

\bibitem{Binosi:2014aea}
D.~Binosi, L.~Chang, J.~Papavassiliou and C.~D. Roberts,
\newblock Phys. Lett. B {\bf 742}, 183 (2015).
%%CITATION = ARXIV:1412.4782;%%

\bibitem{Binosi:2016wcx}
D.~Binosi, L.~Chang, J.~Papavassiliou, S.-X. Qin and C.~D. Roberts,
\newblock Phys. Rev. D {\bf 95}, 031501(R) (2017).
%%CITATION = ARXIV:1609.02568;%%

\bibitem{Binosi:2016nme}
D.~Binosi, C.~Mezrag, J.~Papavassiliou, C.~D. Roberts and
  J.~Rodr{\'i}guez-Quintero,
\newblock Phys. Rev. D {\bf 96}, 054026 (2017).
%%CITATION = ARXIV:1612.04835;%%

\bibitem{Rodriguez-Quintero:2018wma}
J.~Rodr{\'{\i}}guez-Quintero, D.~Binosi, C.~Mezrag, J.~Papavassiliou and C.~D.
  Roberts,
\newblock Few Body Syst. {\bf 59}, 121 (2018).
%%CITATION = ARXIV:1801.10164;%%

\bibitem{Cahill:1987qr}
R.~T. Cahill, C.~D. Roberts and J.~Praschifka,
\newblock Phys. Rev. D {\bf 36}, 2804 (1987).
%%CITATION = PHRVA,D36,2804;%%

\bibitem{Maris:2002yu}
P.~Maris,
\newblock Few Body Syst. {\bf 32}, 41 (2002).
%%CITATION = NUCL-TH/0204020;%%

\bibitem{Maris:2004bp}
P.~Maris,
\newblock Few Body Syst. {\bf 35}, 117 (2004).
%%CITATION = NUCL-TH/0409008;%%

\bibitem{Cahill:1988dx}
R.~T. Cahill, C.~D. Roberts and J.~Praschifka,
\newblock Austral. J. Phys. {\bf 42}, 129 (1989).
%%CITATION = AUJPA,42,129;%%

\bibitem{Burden:1988dt}
C.~J. Burden, R.~T. Cahill and J.~Praschifka,
\newblock Austral. J. Phys. {\bf 42}, 147 (1989).
%%CITATION = AUJPA,42,147;%%

\bibitem{Cahill:1988zi}
R.~T. Cahill,
\newblock Austral. J. Phys. {\bf 42}, 171 (1989).
%%CITATION = AUJPA,42,171;%%

\bibitem{Reinhardt:1989rw}
H.~Reinhardt,
\newblock Phys. Lett. B {\bf 244}, 316 (1990).
%%CITATION = PHLTA,B244,316;%%

\bibitem{Efimov:1990uz}
G.~V. Efimov, M.~A. Ivanov and V.~E. Lyubovitskij,
\newblock Z. Phys. C {\bf 47}, 583 (1990).
%%CITATION = ZEPYA,C47,583;%%

\bibitem{Eichmann:2009qa}
G.~Eichmann, R.~Alkofer, A.~Krassnigg and D.~Nicmorus,
\newblock Phys. Rev. Lett. {\bf 104}, 201601 (2010).
%%CITATION = ARXIV:0912.2246;%%

\bibitem{Cates:2011pz}
G.~Cates, C.~de~Jager, S.~Riordan and B.~Wojtsekhowski,
\newblock Phys. Rev. Lett. {\bf 106}, 252003 (2011).
%%CITATION = ARXIV:1103.1808;%%

\bibitem{Roberts:2013mja}
C.~D. Roberts, R.~J. Holt and S.~M. Schmidt,
\newblock Phys. Lett. B {\bf 727}, 249 (2013).
%%CITATION = ARXIV:1308.1236;%%

\bibitem{Segovia:2015ufa}
J.~Segovia, C.~D. Roberts and S.~M. Schmidt,
\newblock Phys. Lett. B {\bf 750}, 100 (2015).
%%CITATION = ARXIV:1506.05112;%%

\bibitem{Eichmann:2016jqx}
G.~Eichmann,
\newblock Few Body Syst. {\bf 57}, 965 (2016).
%%CITATION = ARXIV:1602.03462;%%

\bibitem{Eichmann:2016hgl}
G.~Eichmann, C.~S. Fischer and H.~Sanchis-Alepuz,
\newblock Phys. Rev. D {\bf 94}, 094033 (2016).
%%CITATION = ARXIV:1607.05748;%%

\bibitem{Eichmann:2016nsu}
G.~Eichmann,
\newblock Few Body Syst. {\bf 58}, 81 (2017).
%%CITATION = ARXIV:1611.10118;%%

\bibitem{Lu:2017cln}
Y.~Lu {\em et~al.},
\newblock Phys. Rev. C {\bf 96}, 015208 (2017).
%%CITATION = ARXIV:1705.03988;%%

\bibitem{Chen:2017pse}
C.~Chen {\em et~al.},
\newblock Phys. Rev. D {\bf 97}, 034016 (2018).
%%CITATION = ARXIV:1711.03142;%%

\bibitem{Mezrag:2017znp}
C.~Mezrag, J.~Segovia, L.~Chang and C.~D. Roberts,
\newblock Phys. Lett. B {\bf 783}, 263 (2018).
%%CITATION = ARXIV:1711.09101;%%

\bibitem{Chen:2018nsg}
C.~Chen {\em et~al.},
\newblock Phys. Rev. D {\bf 99}, 034013 (2019).
%%CITATION = ARXIV:1811.08440;%%

\bibitem{Anselmino:1992vg}
M.~Anselmino, E.~Predazzi, S.~Ekelin, S.~Fredriksson and D.~B. Lichtenberg,
\newblock Rev. Mod. Phys. {\bf 65}, 1199 (1993).
%%CITATION = RMPHA,65,1199;%%

\bibitem{Edwards:2011jj}
R.~G. Edwards, J.~J. Dudek, D.~G. Richards and S.~J. Wallace,
\newblock Phys. Rev. D {\bf 84}, 074508 (2011).
%%CITATION = ARXIV:1104.5152;%%

\bibitem{Roberts:2015lja}
C.~D. Roberts,
\newblock J. Phys. Conf. Ser. {\bf 706}, 022003 (2016).
%%CITATION = ARXIV:1509.02925;%%

\bibitem{Horn:2016rip}
T.~Horn and C.~D. Roberts,
\newblock J. Phys. G. {\bf 43}, 073001 (2016).
%%CITATION = ARXIV:1602.04016;%%

\bibitem{Eichmann:2016yit}
G.~Eichmann, H.~Sanchis-Alepuz, R.~Williams, R.~Alkofer and C.~S. Fischer,
\newblock Prog. Part. Nucl. Phys. {\bf 91}, 1 (2016).
%%CITATION = ARXIV:1606.09602;%%

\bibitem{Arpack}
R.~B.~Lehoucq, D.~C.~Sorensen and C.~Yang, {\em ARPACK Users' Guide: Solution
  of Large-Scale Eigenvalue Problems with Implicitly Restarted Arnoldi
  Methods\/} (Society for Industrial \& Applied Mathematics, 1998).

\bibitem{Maris:1997tm}
P.~Maris and C.~D. Roberts,
\newblock Phys. Rev. C {\bf 56}, 3369 (1997).
%%CITATION = NUCL-TH/9708029;%%

\bibitem{Maris:1999nt}
P.~Maris and P.~C. Tandy,
\newblock Phys. Rev. C {\bf 60}, 055214 (1999).
%%CITATION = NUCL-TH/9905056;%%

\bibitem{Segovia:2015hra}
J.~Segovia {\em et~al.},
\newblock Phys. Rev. Lett. {\bf 115}, 171801 (2015).
%%CITATION = ARXIV:1504.04386;%%

\bibitem{Qin:2018dqp}
S.-X. Qin, C.~D. Roberts and S.~M. Schmidt,
\newblock Phys. Rev. D {\bf 97}, 114017 (2018).
%%CITATION = ARXIV:1801.09697;%%

\bibitem{Eichmann:2008ae}
G.~Eichmann, R.~Alkofer, I.~C. Clo{\"e}t, A.~Krassnigg and C.~D. Roberts,
\newblock Phys. Rev. C {\bf 77}, 042202(R) (2008).
%%CITATION = 0802.1948;%%

\bibitem{Eichmann:2008ef}
G.~Eichmann, I.~C. Clo{\"e}t, R.~Alkofer, A.~Krassnigg and C.~D. Roberts,
\newblock Phys. Rev. C {\bf 79}, 012202(R) (2009).
%%CITATION = 0810.1222;%%

\bibitem{Kamano:2018sfb}
H.~Kamano,
\newblock Few Body Syst. {\bf 59}, 24 (2018).
%%CITATION = FBSYE,59,24;%%

\bibitem{Doring:2018kue}
M.~D{\"o}ring,
\newblock Few Body Syst. {\bf 59}, 140 (2018).
%%CITATION = FBSYE,59,140;%%

\bibitem{Ishii:1998tw}
N.~Ishii,
\newblock Phys. Lett. B {\bf 431}, 1 (1998).
%%CITATION = PHLTA,B431,1;%%

\bibitem{Hecht:2002ej}
M.~B. Hecht {\em et~al.},
\newblock Phys. Rev. C {\bf 65}, 055204 (2002).
%%CITATION = NUCL-TH/0201084;%%

\bibitem{JuliaDiaz:2007kz}
B.~Julia-Diaz, T.~S.~H. Lee, A.~Matsuyama and T.~Sato,
\newblock Phys. Rev. C {\bf 76}, 065201 (2007).
%%CITATION = ARXIV:0704.1615;%%

\bibitem{JuliaDiaz:2007fa}
B.~Julia-Diaz, T.~S.~H. Lee, A.~Matsuyama, T.~Sato and L.~C. Smith,
\newblock Phys. Rev. C {\bf 77}, 045205 (2008).
%%CITATION = ARXIV:0712.2283;%%

\bibitem{Suzuki:2009nj}
N.~Suzuki {\em et~al.},
\newblock Phys. Rev. Lett. {\bf 104}, 042302 (2010).
%%CITATION = ARXIV:0909.1356;%%

\bibitem{Qin:2019hgk}
S.-X. Qin, C.~D. Roberts and S.~M. Schmidt,
\newblock {Few Body Syst. (\emph{in press})}  (arXiv:1902 [nucl-th]),
\newblock {\emph{Spectrum of light- and heavy-baryons}}.
%%CITATION = ARXIV:1902.00026;%%

\bibitem{Yin:2019bxe}
P.-L. Yin {\em et~al.},
\newblock (arXiv:1903.00160 [nucl-th]),
\newblock {\emph{Masses of ground-state mesons and baryons, including those
  with heavy quarks}}.
%%CITATION = ARXIV:1903.00160;%%

\bibitem{Jones:1972ky}
H.~F. Jones and M.~D. Scadron,
\newblock Annals Phys. {\bf 81}, 1 (1973).
%%CITATION = APNYA,81,1;%%

\bibitem{Lorce:2009bs}
C.~Lorce,
\newblock Phys. Rev. D {\bf 79}, 113011 (2009).
%%CITATION = ARXIV:0901.4200;%%

\bibitem{Oettel:1999gc}
M.~Oettel, M.~Pichowsky and L.~von Smekal,
\newblock Eur. Phys. J. A {\bf 8}, 251 (2000).
%%CITATION = NUCL-TH/9909082;%%

\bibitem{Sato:2000jf}
T.~Sato and T.~S.~H. Lee,
\newblock Phys. Rev. C {\bf 63}, 055201 (2001).
%%CITATION = NUCL-TH/0010025;%%

\bibitem{Eichmann:2011vu}
G.~Eichmann,
\newblock Phys. Rev. D {\bf 84}, 014014 (2011).
%%CITATION = 1104.4505;%%

\bibitem{Aznauryan:2009mx}
I.~Aznauryan {\em et~al.},
\newblock Phys. Rev. C {\bf 80}, 055203 (2009).

\bibitem{Beck:1999ge}
R.~Beck {\em et~al.},
\newblock Phys. Rev. C {\bf 61}, 035204 (2000).
%%CITATION = NUCL-EX/9908017;%%

\bibitem{Pospischil:2000ad}
T.~Pospischil {\em et~al.},
\newblock Phys. Rev. Lett. {\bf 86}, 2959 (2001).
%%CITATION = NUCL-EX/0010020;%%

\bibitem{Blanpied:2001ae}
G.~Blanpied {\em et~al.},
\newblock Phys. Rev. C {\bf 64}, 025203 (2001).
%%CITATION = PHRVA,C64,025203;%%

\bibitem{Sparveris:2004jn}
N.~Sparveris {\em et~al.},
\newblock Phys. Rev. Lett. {\bf 94}, 022003 (2005).
%%CITATION = NUCL-EX/0408003;%%

\bibitem{Stave:2008aa}
S.~Stave {\em et~al.},
\newblock Phys. Rev. C {\bf 78}, 025209 (2008).
%%CITATION = ARXIV:0803.2476;%%

\bibitem{Maris:1998hc}
P.~Maris and C.~D. Roberts,
\newblock Phys. Rev. C {\bf 58}, 3659 (1998).
%%CITATION = NUCL-TH/9804062;%%

\bibitem{Capstick:1994ne}
S.~Capstick and B.~D. Keister,
\newblock Phys. Rev. D {\bf 51}, 3598 (1995).
%%CITATION = NUCL-TH/9411016;%%

\bibitem{Aznauryan:2015zta}
I.~G. Aznauryan and V.~D. Burkert,
\newblock Phys. Rev. C {\bf 92}, 035211 (2015).
%%CITATION = ARXIV:1506.03183;%%

\bibitem{Aznauryan:2016wwm}
I.~G. Aznauryan and V.~D. Burkert,
\newblock (2016),
\newblock {\emph{Configuration mixings and light-front relativistic quark model
  predictions for the electroexcitation of the \mbox{$\Delta(1232)3/2^+$},
  \mbox{$N(1440)1/2^+$}, and \mbox{$\Delta(1600)3/2^+$}}, arXiv:1603.06692
  [hep-ph]}.
%%CITATION = ARXIV:1603.06692;%%

\bibitem{Idilbi:2003wj}
A.~Idilbi, X.-d. Ji and J.-P. Ma,
\newblock Phys. Rev. D {\bf 69}, 014006 (2004).
%%CITATION = HEP-PH/0308018;%%

\bibitem{NAP13438}
\mbox{National Research Council},
\newblock {\em Nuclear Physics: Exploring the Heart of Matter} (The National
  Academies Press, Washington, DC, 2013).

\bibitem{Brodsky:2015aia}
S.~J. Brodsky {\em et~al.},
\newblock (aXiv:1502.05728 [hep-ph]),
\newblock {\emph{QCD and Hadron Physics}}.
%%CITATION = ARXIV:1502.05728;%%

\bibitem{Braun:2009jy}
V.~Braun {\em et~al.},
\newblock Phys. Rev. Lett. {\bf 103}, 072001 (2009).
%%CITATION = ARXIV:0902.3087;%%

\bibitem{Capstick:1992xn}
S.~Capstick,
\newblock Phys. Rev. D {\bf 46}, 1965 (1992).
%%CITATION = PHRVA,D46,1965;%%

\bibitem{Melde:2008yr}
T.~Melde, W.~Plessas and B.~Sengl,
\newblock Phys. Rev. D {\bf 77}, 114002 (2008).
%%CITATION = ARXIV:0806.1454;%%

\bibitem{deTeramond:2014asa}
G.~F. de~Teramond, H.~G. Dosch and S.~J. Brodsky,
\newblock Phys. Rev. D {\bf 91}, 045040 (2015).
%%CITATION = ARXIV:1411.5243;%%

\bibitem{Park:2014yea}
K.~Park {\em et~al.},
\newblock Phys. Rev. C {\bf 91}, 045203 (2015).
%%CITATION = ARXIV:1412.0274;%%

\bibitem{Mokeev:2012vsa}
V.~I. Mokeev {\em et~al.},
\newblock Phys. Rev. C {\bf 86}, 035203 (2012).
%%CITATION = ARXIV:1205.3948;%%

\bibitem{Mokeev:2015lda}
V.~I. Mokeev {\em et~al.},
\newblock Phys. Rev. C {\bf 93}, 025206 (2016).
%%CITATION = ARXIV:1509.05460;%%

\bibitem{Dugger:2009pn}
M.~Dugger {\em et~al.},
\newblock Phys. Rev. C {\bf 79}, 065206 (2009).
%%CITATION = 0903.1110;%%

\end{thebibliography}

\end{document}